\documentclass[a4paper,11pt]{article}
\pdfoutput=1 
\usepackage{jcappub} 

\usepackage[T1]{fontenc} 
\usepackage[applemac]{inputenc}

\usepackage{amsmath}
\usepackage{amssymb}
\usepackage[numbers]{natbib}
\usepackage{graphicx}
\usepackage{multirow}
\usepackage{enumerate}
\usepackage[normalem]{ulem}
\usepackage{empheq}
\usepackage{subcaption}

\newcommand{\rd}{\mathrm{d}}
\newcommand{\re}{\mathrm{e}}
\newcommand{\Oin}{\Omega^{\text{(L)}}}
\newcommand{\Oout}{\Omega^{\text{(R)}}}
\newcommand{\CR}{\mathcal{R}}

\newcommand{\jump}[1]{\left[#1\right]}
\newcommand{\mean}[1]{\langle#1\rangle}

\renewcommand{\vec}{\boldsymbol}

\title{\boldmath Cosmology on a cosmic ring}
 
\author[a,c]{Florian Niedermann}
\author[a,c]{and  Robert Schneider}
\affiliation[a]{Arnold Sommerfeld Center for Theoretical Physics, Ludwig-Maximilians-Universit\"at, Theresienstra{\ss}e 37, 80333 Munich, Germany}
\affiliation[c]{Excellence Cluster Universe, Boltzmannstra{\ss}e 2, 85748 Garching, Germany\\
~\\}
 
\emailAdd{florian.niedermann@physik.lmu.de}
\emailAdd{robert.bob.schneider@physik.uni-muenchen.de}

\abstract{

We derive the modified Friedmann equations for a generalization of the Dvali-Gabadadze-Porrati (DGP) model in which the brane has one additional compact dimension. The main new feature is the emission of gravitational waves into the bulk. We study two classes of solutions: First, if the compact dimension is stabilized, the waves vanish and one exactly recovers DGP cosmology. However, a stabilization by means of physical matter is not possible for a tension-dominated brane, thus implying a late time modification of 4D cosmology different from DGP. Second, for a freely expanding compact direction, we find exact attractor solutions with zero 4D Hubble parameter despite the presence of a 4D cosmological constant. The model hence constitutes an explicit example of dynamical degravitation at the full nonlinear level. Without stabilization, however, there is no 4D regime and the model is ruled out observationally, as we demonstrate explicitly by comparing to supernova data.

}

\begin{document}

\maketitle
\flushbottom

\section{Introduction}

Cosmological observations probing the expansion history of our universe unambiguously indicate a recent transition to an accelerated expansion stage in accordance with a vacuum dominated energy budget \cite{Perlmutter:1998np,Riess:1998cb}. However, from a theoretical point of view there is still a lack of fundamental understanding of this so called dark energy. Naive evaluations within in the effective field theory approach lead to expectations for the vacuum energy density that are at least 35 orders of magnitude above the value inferred from the observed curvature scale. This discrepancy is set by a low energy standard model parameter, the electron mass, and gets worse as the entire 
standard model particle spectrum gets successively integrated in. Since this challenge is posed at energies far below the electroweak scale, it is unlikely to be resolved solely by new UV physics unless some unexpected conspiracy between high and low energy physics is at work. 

This so called cosmological constant problem\footnote{See \cite{Weinberg:1988cp} for a seminal work and \cite{Burgess:2011va} for a more recent discussion.} requires 4D Einstein gravity to be at work. Only then Einstein's equations assign a definite energy density to the observed curvature scale. Therefore, a promising way towards a possible solution seems to lie within the gravitational sector itself. The idea is to weaken the gravitational force at large distances and thereby making it insensitive to a cosmological constant. This approach was termed {\it degravitation} and has been pursued within different models (see in particular~\cite{Dvali:2007kt} but also~\cite{Dvali:2002pe, Dvali:2002fz, ArkaniHamed:2002fu, deRham:2007rw, Burgess:2011va}).

Infrared modifications of gravity are also interesting for another reason. As they generically lead to a late time modification of the standard Friedmann evolution, they allow to challenge General Relativity (GR) on cosmological scales. In other words, developing consistent competitor theories enables us to quantify the phenomenological success of GR by comparing predictions of both models with observations. In this work we are going to present a model which turns out to be relevant with respect to both motivations: We will find a solution that features a new degravitation mechanism and another solution which leads to a potentially viable late time modification of GR.

There are different ways to modify GR in the infrared: One class of models attempts to give a mass to the graviton. This approach has been studied extensively on a linear level and lately been realized on a non-linear level \cite{deRham:2010kj} (for reviews see~\cite{Hinterbichler:2011tt,deRham:2014zqa}). The class of models we are focussing on relies on introducing additional spatial dimensions.  We are interested in those models with at least one infinite extra dimension\footnote{That property turns out to be important to circumvent Weinberg's no-go result.}, in which all matter fields are localized on a brane, but gravity is intrinsically higher dimensional. To recover 4D gravity at short distances, a brane-localized four dimensional Einstein-Hilbert term is included. Therefore, these models are referred to as {\it brane induced gravity} (BIG) \cite{Dvali:2000hr,Dvali:2000xg}.

Within this higher dimensional picture, the degravitation mechanism can be easily understood in terms of intrinsic and extrinsic curvature. Since the vacuum energy is localized on the brane, it breaks the higher dimensional Poincar\'{e} invariance and, in principle, allows for vacuum solutions with a flat brane and a curved bulk geometry, i.e.\ vanishing intrinsic and non-vanishing extrinsic curvature. Thus, for a 4D observer it would appear as if gravity was insensitive to a vacuum energy or, equivalently, a cosmological constant \cite{Rubakov:1983bz,Charmousis:2001hg}. As promising as this mechanism looks, as difficult it turned out to be implemented.

In the case of one infinite extra dimension, this is the Dvali-Gabadadze-Porrati (DGP) model \cite{Dvali:2000hr}. Its cosmological solutions were derived in \cite{Deffayet:2000uy} and have been studied extensively. Although it modifies the standard Friedmann cosmology, it does not lead to degravitation. Moreover, one of the two branches of solutions turned out to suffer from perturbative ghost instabilities \cite{Luty:2003vm,Nicolis:2004qq,Koyama:2005tx,Charmousis:2006pn,Gregory:2007xy,Gorbunov:2005zk}. The other (healthy) branch is clearly not favored as compared to $ \Lambda $CDM by cosmological observations~\cite{Lombriser:2009}.

The failure of the 5D model as a phenomenologically viable modification motivates the study of its higher dimensional generalizations. We will limit ourselves to the case of six dimensions. { \it The idea of the present paper is to supplement the one infinite extra dimension of the DGP model with another compact brane dimension.} Thus, the brane is a codimension one object and, in an embedding picture, corresponds to a ring that is wrapped around an infinite cylinder. We will investigate the cosmology of this model by deriving a set of modified Friedmann equations which describe the curvature evolution on the brane. To that end, we assume that the metric and the localized energy momentum tensor possess 4D FRW symmetries.
A new feature of this model is the existence of bulk gravitational waves. (In 5D the situation is different since there is a version of Birkhoff's theorem \cite{Taub:1951,Hofmann:2013zea} which applies to planar symmetry and implies a static bulk.) Since the bulk is assumed to be empty, the only emitter of gravitational waves is the brane itself, and the physical requirement of having no incoming bulk waves has to be imposed. 

Another common generalization to 6D consists in introducing two infinite extra dimensions and making the brane a codimension two object \cite{Dvali:2000xg,Kaloper:2007ap}. The (regularized) geometry corresponds to a (capped) cone and is thus topologically distinct from the one we are investigating. The codimension two model bears certain technical difficulties due to its 2D radial symmetry. In particular, it is not possible to exclude incoming waves by means of a {\it local} criterion. In fact, it is known that such an ``outgoing wave condition'' for cylindrical waves is essentially non-local in time \cite{Givoli:1991}. This makes it impossible to derive a closed set of modified Friedmann equations that are local in time and requires numerical methods to further investigate the model. This has been achieved recently in \cite{Niedermann:2014bqa}. The advantage of the setup at hand is that there is a local wave criterion for plane waves which makes it possible to derive a closed system of modified Friedmann equations and in turn allows to infer analytic properties of the solutions.

Let us emphasize that the main motivation for the 6D model of this work is that it serves as the simplest prototype for models which have both infinite bulk as well as compact brane extra dimensions\footnote{One major motivation to study extra-dimensional models with more than one extra dimension (apart from the fact that they give rise to a huge class of consistent GR modifications) is of course provided by string theory, which is only consistent in 10 or 11 dimensions. } and which --- due to the induced Einstein-Hilbert term and the compactness of the additional brane dimensions --- can have a 4D gravity regime. As such, it is already capable of introducing an important physical feature which is expected to generically occur in all other (higher dimensional) setups, namely the emission of gravitational waves into bulk. Yet, it is still simple enough to allow for analytic solutions, which greatly simplifies the discussion of its physical implications.
Furthermore, the cosmological solutions that we find turn out to have some very intriguing properties, giving (at least) two \textit{a posteriori} motivations:
(i)~The model provides a new degravitation mechanism at the full nonlinear level, which --- albeit being ruled out observationally in this particular case --- can be considered a proof of principle. This might also serve as an inspiration to come up with other similar (and hopefully phenomenologically viable) models that are able to degravitate the cosmological constant.
(ii)~It could yield a potentially interesting late time-modification of the DGP cosmology. As will be explained below, this is due to a breakdown of modulus stabilization of the compact extra dimension. Therefore, this mechanism might be relevant for all other higher dimensional theories with stabilized compact extra dimension, and it might open the door to a novel class of consistent GR competitor theories which could be put to the test on cosmologically relevant scales.

In Section \ref{sec:model} the model is introduced. Special emphasis is placed on the cylindrical geometry which distinguishes it from the codimension two model. Since we are interested in curvature quantities that can be inferred by an on-brane observer, we attempt to find a closed system of differential equations for the induced metric on the brane. To that end, we derive the junction conditions at the brane in Section~\ref{sec:matching}. However, these equations do not yet constitute a closed system. As mentioned before, this is related to the existence of bulk gravitational waves: in general, incoming waves will affect the curvature evolution on the brane. Therefore, we have to implement a differential condition that excludes incoming waves in accordance with a source-free bulk. This can be done in a coordinate independent way by employing the Weyl tensor. This is done in Section \ref{sec:closed_system}, yielding a closed system of equations. We further exclude any static curvature component in the bulk which eliminates one integration constant and in turn yields the modified Friedmann equations in their final form.

In Section \ref{sec:solution} we consider two different classes of solutions. Of particular physical interest are solutions for which the compact dimension is stabilized. This can always be achieved by some underlying stabilization mechanism which is effectively implemented by tuning the azimuthal pressure $p_\phi$. In this case we exactly reproduce the DGP equations. Therefore, these solutions correspond to the trivial embedding of the DGP solutions in the higher dimensional space. Moreover, bulk waves are absent for these solutions which is in accordance with the DGP result for which the brane is embedded in a static bulk. However, we find that for a tension dominated brane source, the stabilization can no longer be realized by means of physical matter. More specifically, the corresponding equation of state parameter would need to be $w_{\phi}\leq-4/3$ which signals a violation of the Null Energy Condition. Therefore, it is argued that the size of the compact dimension starts to evolve for late times when the brane enters the regime of tension domination and all other matter components have thinned out sufficiently. In general, the non-trivial dynamics of the compact direction leads to a modification of the standard 4D Friedmann evolution which is different from the DGP case. A more quantitative statement requires a microscopic description of the stabilization mechanism and is left for future work.

Of particular conceptual interest are solutions with an expanding or collapsing compact dimension. As an exemplary case, we discuss $p_{\phi}=0$. For a pure tension source, we find analytic solutions in which the Hubble parameter, describing the evolution of the three non-compact brane directions, vanishes. Thus, an effective 4D brane observer will measure a vanishing curvature despite the presence of a 4D spacetime homogeneous energy source. We also solve the equations numerically for a non-vanishing initial Hubble parameter and find that the degravitating solution is an attractor. This is an important result of this work, since it constitutes a {\it stable, dynamical example of the degravitation proposal at the fully non-linear level}. Another example was recently obtained in the codimension two setup \cite{Niedermann:2014bqa}\footnote{There, it was necessary to solve for the full bulk geometry numerically, whereas here we are able to derive local on-brane evolution equations. These ordinary differential equations are technically much easier to analyze.}. The physical mechanism at work, however, differs in the two scenarios: In codimension two, the effect of the cosmological constant $ \Lambda $ is to curve the transverse spatial dimensions (into a cone), whereas here, it is the temporal curvature (i.e.\ expansion) in the compact extra dimension which absorbs $ \Lambda $. Furthermore, it is argued that the mechanism relies on the existence of an infinite extra dimension and as a consequence allows to evade Weinberg's famous no-go theorem~\cite{Weinberg:1988cp}. However, it is not yet clear whether the choice  $p_{\phi}=0$ implies another fine-tuning.

Since the degravitating solution is not compatible with a stabilized compact dimension, there are observable effects related to the corresponding size modulus. The existence of such a scalar degree of freedom during all cosmological epochs questions the phenomenological viability of the model since, generically, it leads to deviations from a standard 4D behavior. We perform a supernova fit which clearly shows that the degravitating solution with $p_{\phi}=0$ is ruled out. 

We conclude in Section \ref{sec:conclusion}. The full bulk solution is presented in the appendix.
We adopt the following notational conventions: capital Latin indices $ A, B, \dots $ denote six-dimensional, small Latin indices $ a, b, \ldots $ five-dimensional  spacetime indices. Small Latin indices $ i, j, \ldots $ run over the three non-compact spatial on-brane dimensions.
The spacetime dimensionality $ d $ of some quantity $ Q $ is sometimes made explicit by a writing $ Q^{(d)} $.
Our sign conventions are ``$ +++ $'' as defined (and adopted) in~\cite{Misner}.
We work in units in which $ c = \hbar = 1 $.

\section{The model}
\label{sec:model}

The action of our model is the sum of three terms,
\begin{align} 
	\label{ActionBIG}
	\mathcal{S}=
	\mathcal{S}_{\rm EH}+
	\mathcal{S}_{\rm BIG}+
	\mathcal{S}_{\rm m}\;,
\end{align}
where
\begin{align}
	\mathcal{S}_{\rm EH}=
	M_{6}^{4}\int \rd^6 X\, \sqrt{-g} \CR^{(6)}
\end{align}
is the Einstein-Hilbert term describing gravity in a six-dimensional bulk spacetime. The second term
\begin{align}
	\label{S_BIG_delta_0}
	\mathcal{S}_{\rm BIG } = M_5^3 \int \rd^5 x\,\sqrt{-h}\, \CR^{(5)}
\end{align}
is normally referred to as the \textit{brane induced gravity} (BIG) term. In an effective field theory context it arises due to some heavy fields that are confined on a lower-dimensional hypersurface called the brane. These fields have been integrated out in the low energy regime of the theory and effectively modify the graviton dynamics. 

\begin{figure}[htb]
	\begin{subfigure}{0.48\textwidth}
	\centering
	\includegraphics[width=\textwidth]{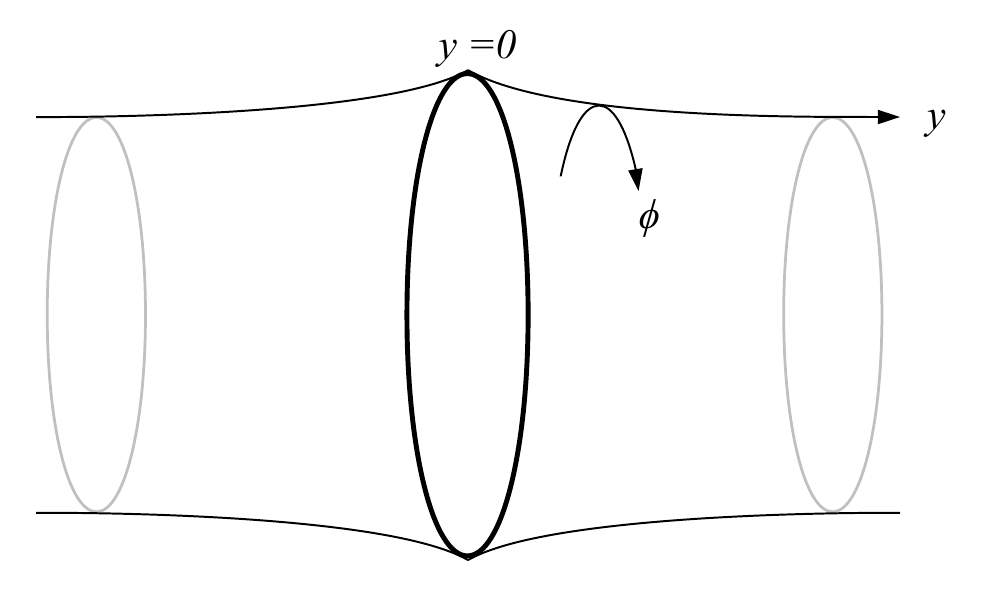}
	\caption{\textit{Cylindrical geometry:} there is no symmetry axis, and it is natural to assume reflection symmetry around $ y=0 $.}
	\label{fig:cylinder}
	\end{subfigure}
	\hfill
	\begin{subfigure}{0.48\textwidth}
	\centering
	\includegraphics[width=\textwidth]{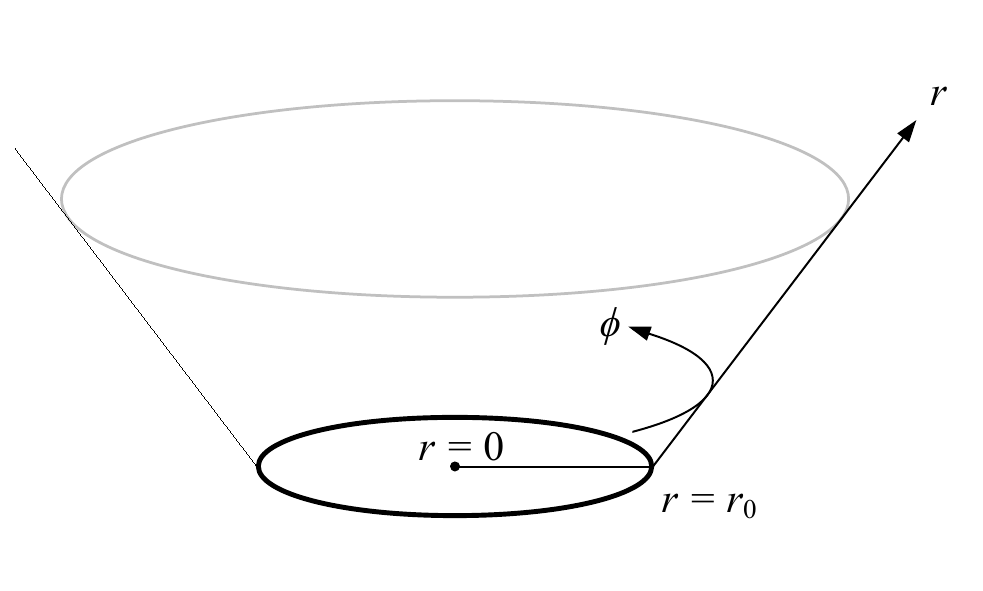}
	\caption{\textit{Radial geometry:} $ r=0 $ is the symmetry axis and the brane can create a defect angle in the exterior. There is no reflection symmetry around $ r = r_0 $ in this case.}
	\label{fig:capped_cone}
	\end{subfigure}
	\caption{Illustration of two possible bulk geometries that could locally be described by the same metric ansatz, but are topologically different. In this work, we are considering the cylindrical topology (a), see also Section \ref{sec:distinction_codim_2}. Only the two extra dimensions are drawn, embedded into a fictitious three-dimensional space. The 5D brane is located at $ y=0 $ and $ r = r_0 $, respectively.}
	\label{fig:geometry}
\end{figure}

We are considering a model in which the brane has five dimensions --- the four infinite spacetime dimensions we see in low-energy physics, plus one compact (spatial) dimension of microscopic length $ 2\pi R $ --- and the bulk has one additional infinite dimension, cf.\ Figure~\ref{fig:cylinder}. This should not be confused with a model with two infinite extra dimensions, with the codimension two brane regularized by blowing it up to a circle of radius $ r_0 $, cf.\ Figure~\ref{fig:capped_cone}, which is topologically different. This model is discussed elsewhere \cite{Niedermann:2014bqa}.

By dimensional reduction, the mass scale $ M_5 $ is related to the four-dimensional Planck mass by $ M_4^2 = 2\pi R M_5^3 $. The last term in \eqref{ActionBIG} denotes the matter part of the action that is strictly localized on the brane.
In the expressions above, $ g_{AB} $ denotes the six-dimensional bulk metric and $ h_{ab} $ is the five-dimensional metric induced on the brane.

\subsection{Gaussian normal coordinates}

We are looking for a classical background geometry which is independent of the compact extra-dimensional coordinate $ \phi \in [0, 2\pi) $, as well as homogeneous and isotropic (and for simplicity spatially flat) in the three spatial brane dimensions, labeled by the coordinates $ x^i $.
The most general metric ansatz can thus be written as
\begin{align}
	\label{eq:general_metric}
	\rd s^2 = -\re^{2n(t, y)}\rd t^2 + \re^{2f(t, y)} \rd t \, \rd y + \re^{2a(t, y)} \delta_{ij} \rd x^i \rd x^j + \re^{2b(t, y)} \,\rd y^2 + \re^{2c(t, y)}\,   \rd \phi^2\;,
\end{align}
where $ y $ is the coordinate corresponding to the infinite extra dimension.
The brane's position in extra-space will then in general be given by a worldline of the form $ (t, y_0(t)) $. To make the junction conditions as simple as possible, however, it is convenient to introduce \textit{Gaussian normal coordinates} \citep[see, e.g.][Appendix D]{Carroll}: the new radial coordinate (which we will call $ y $ again) is defined to be the proper distance along space-like geodesics perpendicular to the brane. This is always possible (at least locally; the geodesics can in principle cross at some finite distance away from the brane, which would lead to a coordinate singularity at that point), and implies that:
\begin{enumerate}[(i)]
\item the brane is located at the fixed coordinate $ y = 0 $;
\item the metric now takes the form\footnote{By a slight abuse of notation, we use the same names for the coordinates and functions as before.}
\begin{subequations}
\label{eq:met_GNC}
\begin{align}
	\rd s^2 & = \rd y^2 + \sum_{x^A \neq y} g_{AB} \, \rd x^A \rd x^B \\
	& = -\re^{2n(t, y)}\rd t^2 + \re^{2a(t, y)} \delta_{ij} \rd x^i \rd x^j + \rd y^2 + \re^{2c(t, y)} \rd \phi^2\;.
\end{align}
\end{subequations}
\end{enumerate}

In Gaussian normal coordinates, implementing Israel's junction conditions is equivalent to inserting delta function terms into Einstein's field equations. The complete set of equations of motion can thus be written as
\begin{equation}
\label{eq:einstein}
	M_6^4 G^{(6)}_{AB} = \delta(y) \delta^a_A \delta^b_B \left(T_{ab} - M_5^3 \, G^{(5)}_{ab} \right) \, .
\end{equation}
Here $ T_{ab} $ is the five-dimensional surface energy momentum tensor localized on the brane. The most general form compatible with the symmetries of our set-up that it can take is
\begin{equation}
\label{eq:EMT}
	T^{a}_{\hphantom{a}b} = \mathrm{diag}\left( -\rho,\, p,\, p,\, p,\, p_{\phi} \right) \,.
\end{equation}
The pressure component in angular direction $ \phi $ can be used to stabilize the brane circumference. We will come back to this point later.

\section{On-brane equations}
\label{sec:matching}
For an observer living on the brane, only the intrinsic brane curvature can be directly inferred from gravitational or cosmological measurements. Thus, it would be extremely useful to derive the modified Friedmann equations stemming from \eqref{eq:einstein} which describe the evolution of the scale factor on the brane, but without having to solve for the full bulk geometry. Therefore, one of the main purposes of this paper consists in deriving a closed system for the variables $ a_0(t) := a(t,0)$, $c_0(t) := c(t,0)$ and $ n_0(t) := n(t,0) $ which are evaluated at the position of the brane $y=0$.

In fact, we can already make our lives a bit easier by employing the residual gauge freedom of \eqref{eq:met_GNC}, i.e.\ a redefinition of the time coordinate $ t $ at the position of the brane, to set $ n_0 \equiv 0 $.
As a result, the induced metric that a brane observer can measure is given by
\begin{equation}
\label{eq:ind_met}
	\rd s_{(5)}^2 = -\rd t^2 + \re^{2a_0(t)} \delta_{ij} \rd x^i \rd x^j + \re^{2c_0(t)} \rd \phi^2 \, ,
\end{equation}
and we would like to determine the two unknown functions $ a_0(t) $ and $ c_0(t) $ for any given matter source.

\subsection{Modified Einstein equations}

In order to illustrate the basic properties of the system \eqref{eq:einstein}, let us write down Einstein's field equations explicitly. Inserting \eqref{eq:met_GNC} into \eqref{eq:einstein} results in five independent equations for the functions $n(t,y)$, $a(t,y)$ and $c(t,y)$:

\begin{subequations}
\label{full_system}
\begin{align}
\label{tt}(tt) & : \quad \re^{-2 n} \left(-3 \dot{a}^2-3 \dot{a} \dot{c}\right) + 6 {a'}^2+3 a' c'+{c'}^2+3 a''+c''
=- \frac{\delta (y)}{M_6^4}\;  \tilde\rho  \\
\label{ii}(ii) & : \quad \re^{-2 n} \left(-3 \dot{a}^2-2 \dot{a} \dot{c}-\dot{c}^2+2 \dot{a} \dot{n}+\dot{c} \dot{n}-2 \ddot{a}-\ddot{c}\right) \\
& \qquad +3 {a'}^2+2 a' c'+{c'}^2+2 a' n'+c' n'+{n'}^2+2 a''+c''+n''
= \frac{\delta (y)}{ M_6^4}\;    \tilde p  \nonumber\\
\label{pp}(\phi\phi) \!-\!(yy)	
& :\quad \re^{-2 n} \left(3 \dot{a} \dot{c}+\dot{c}^2-\dot{c} \dot{n}+\ddot{c}\right) + 3 {a'}^2-3 a' c'-c' n'+{n'}^2+3 a''+n''
=\frac{\delta (y) }{   M_6^4}\;   \tilde p_{\phi}  \\
\label{yy}(yy) & : \quad \re^{-2 n} \left(-6 \dot{a}^2-3 \dot{a} \dot{c}-\dot{c}^2+3 \dot{a} \dot{n}+\dot{c} \dot{n}-3 \ddot{a}-\ddot{c}\right) +3 {a'}^2+3 a' c'+3 a' n'+c' n' = 0\\
\label{ty}(ty) &: \quad \re^{-2 n} \left(3 a' \dot{a}-3 n' \dot{a}+c' \dot{c}-n' \dot{c}+3 \dot{a}'+\dot{c}'\right) = 0
\end{align}
\end{subequations}
Here a dot is shorthand notation for the partial derivative $ \partial / \partial_t $, and a prime denotes $ \partial / \partial_y $. Furthermore, we  employ a notation in which all the brane induced terms are absorbed in modified expressions for the pressures and the energy density. This is useful because many of the general statements in this paper do not depend on the existence of the brane induced gravity term and would therefore equally apply in a setup where only a perfect fluid source is localized on the brane. Specifically,
\begin{subequations}
\label{eq:rho_p_tilde}
	\begin{align}
		\tilde \rho & := \rho - 3 M_5^3 \left( H_a^2 + H_a H_c \right) \,,\\
		\tilde p& := p + M_5^3 \left( 2\dot H_a + \dot H_c + 3 H_a^2 + H_c^2 + 2 H_a H_c \right) \,,\\
		\tilde p_\phi & := p_\phi+ 3 M_5^3 \left( \dot H_a + 2 H_a^2 \right)\,.
	\end{align}
\end{subequations}
Here we also introduced the two ``Hubble'' parameters associated with the on-brane metric~\eqref{eq:ind_met}, $H_a := \dot a_0 $ and $ H_c := \dot c_0$, measuring the expansion rate in $ x $- and $ \phi $-direction, respectively.

\subsection{From partial to ordinary differential equations}
\label{sec:from_PDE_to_ODE}

In the general case, equations \eqref{full_system} constitute a complicated system of partial differential equations (PDE). However, we are only interested in the on-brane dynamics, i.e.\ in the dynamics of the variables evaluated at $y=0$.  The idea is to extract this information from the system  without knowing the detailed $y$-profile of the metric functions.

But we then have to face the question of how to deal with the $y$-derivatives of the metric functions in \eqref{full_system}. The appearance of the delta functions will introduce a non-regular behavior of the solution in the $y$-direction at the position of the brane. In fact, the only possibility to create the delta functions consists in assuming that the second $y$-derivative terms themselves contain a part that is proportional to a delta function, so that we can write
\begin{subequations}
\begin{align}
	a^{\prime \prime} & = \hat a^{\prime \prime} + \delta(y) \jump{a'} \,, \\
	c^{\prime \prime} & = \hat c^{\prime \prime} + \delta(y)  \jump{c'} \,, \\
	n^{\prime \prime} & = \hat n^{\prime \prime} + \delta(y) \jump{n'}\,,
\end{align}
\end{subequations}
where $\hat a^{\prime \prime} $, $\hat c^{\prime \prime} $ and $\hat n^{\prime \prime} $ do not contain any further delta function contributions. The squared brackets denote the \textit{jump} of the first derivative of the corresponding quantity across the brane:
\begin{equation}\label{def_jump}
	\jump{f} := f(t, 0^+) - f(t, 0^-) \;,
\end{equation}
where $f$ is an arbitrary function of $t$ and $y$.
This allows us to match all delta function parts of the equations  \eqref{full_system}, yielding the following \textit{junction conditions}: 
\begin{subequations}
	\label{junction_cond}
\begin{align}
	(tt): 	&\qquad  3 \jump{a'} + \jump{c'} =- \frac{1}{ M_6^4} \, \tilde \rho\\
	(ii):	&\qquad  2 \jump{a'} + \jump{c'} + \jump{n'} = \frac{1}{M_6^4} \, \tilde p\\
	(\phi\phi): &\qquad 3 \jump{a'} + \jump{n'} = \frac{1}{M_6^4} \, \tilde p_{\phi}
\end{align}
\end{subequations}
The $(yy)$ and $(ty)$-equations do not contain any delta functions. 
Note that this set of equations is equivalent to Israel's junction conditions \cite{Israel:1966, Israel:1967}. The delta-function procedure can be used here because we are working in Gaussian normal coordinates \cite{Stephani}.

Solving \eqref{junction_cond} for the jumps yields
\begin{subequations}
	\label{junction_cond_solved}
	\begin{align}
		\jump{a'} &= \frac{1}{4 M_6^4} \left( \tilde p_{\phi} -\tilde p -\tilde \rho \right)   \label{X}\,, \\
		\jump{c'} &= \frac{1}{4 M_6^4} \left( 3\tilde p - 3\tilde p_{\phi} - \tilde \rho \right)   \label{Y}\,, \\
		\jump{n'} &= \frac{1}{4 M_6^4} \left(  \tilde p_{\phi} +3 \tilde p + 3\tilde \rho \right)   \label{Z}\,.
	\end{align}
\end{subequations}

There is more information that can be extract from the system \eqref{full_system} in a small vicinity of the brane. As the metric functions themselves are allowed to have a kink at the position of the brane, the first $y$-derivatives ($a'$, $b'$ and $n'$), as well as the non-distributional parts of the second $y$-derivatives  ($\hat a'' $, $\hat b'' $ and $\hat n'' $) posses a discontinuity at the position of the brane. It is fully characterized by the jump, as defined in \eqref{def_jump}, as well as the \textit{mean} across the brane:

\begin{equation}\label{def_mean}
	\mean{f} := \frac{f(t,0^+) + f(t, 0^-)}{2} \;.
\end{equation}
Thus, in order to derive further on-brane equations, we simply have to calculate the jump and the mean of each equation in the system \eqref{full_system}. In fact, these are all the on-brane equations that can be extracted from the full set of Einstein equations.
Since there are terms that are quadratic in the discontinuous functions, we use the relations
\begin{align}
	\jump{fg}=\jump{f}\mean{g} + \jump{g}\mean{f} && \text{and} && \mean{fg} = \mean{f}\mean{g} + \frac{1}{4}\jump{f}\jump{g}\;,
\end{align}
which hold for any arbitrary functions $f(y)$ and $g(y)$. By that procedure the five PDEs are reduced to a system of ten ordinary differential equations (ODE) in the variable $t$. Since the $(tt)$, $(ii)$ and $(\phi\phi)$-components of the Einstein equations are second order in $y$-derivatives, they yield six equations which can be used to solve for $\jump{f''}$ and $\mean{f''}$ with $f \in \{a,n,c\}$. The $(yy)$ and $(ty)$-components, which are both first order in  $y$-derivatives, yield the remaining four equations:

\begin{subequations}
\label{ODE_full}
 \begin{align}
	\label{rr_ODE_J}\jump{yy} & : \; 3\jump{a'} \Bigl( 2\mean{a'} + \mean{c'} + \mean{n'} \Bigr) + \jump{c'} \Bigl ( 3\mean{a'} + \mean{n'} \Bigr) + \jump{n'}\Bigl ( 3\mean{a'} + \mean{c'} \Bigr) = 0 \,, \\
	\label{rr_ODE_M}\mean{yy} & : \; \frac{3}{4} \left[a'\right] \Bigl( \jump{a'} + \jump{c'} + \jump{n'} \Bigr) +\frac{1}{4} \jump{c'} \jump{n'} + 3 \mean{a'} \Bigl( \mean{a'} + \mean{c'} + \mean{n'} \Bigr) + \mean{c'} \mean{n'} \nonumber\\
	& \qquad\qquad\qquad\qquad\qquad\qquad\qquad\qquad -6 H_a^2-3 H_a H_c-H_c^2-3 \dot H_a-\dot H_c = 0 \,, \\
	\label{tr_ODE_J}\jump{ty} & : \;  3 H_a \Bigl( \jump{a'} - \jump{n'} \Bigr) + H_c \Bigl(\jump{c'} - \jump{n'} \Bigr) + 3 \dot{\jump{a'}} + \dot{\jump{c'}} = 0 \,, \\
	\label{tr_ODE_M}\mean{ty}	&: \; 3H_a \Bigl( \mean{a'} - \mean{n'} \Bigr) + H_c \Bigl( \mean{c'} - \mean{n'} \Bigr) + 3\dot{\mean{a'}} +\dot{\mean{n'}} = 0
 \end{align}
 \end{subequations}
It is straightforward to check that after inserting \eqref{junction_cond_solved} in equation \eqref{tr_ODE_J}, it becomes
\begin{equation}
\label{eq:en_cons}
	\dot \rho + 3 H_a \left( \rho +  p \right) + H_c \left( \rho +  p_\phi \right) = 0 \,.
\end{equation}
This is nothing but the $ t $-component --- which, due ot the symmetries, is the only non-trivial one --- of the covariant conservation of the energy momentum tensor as defined in \eqref{eq:EMT}: $\nabla^{(5)}_a T^{(5)a}_{\hphantom{(5)a}b}=0$. Note that the brane induced gravity terms dropped out in \eqref{eq:en_cons} due to the Bianchi identity.

Let us now discuss the system \eqref{ODE_full} of ODEs qualitatively. Since $\jump{a'}$, $\jump{c'}$ and $\jump{n'}$ can be expressed by virtue of \eqref{eq:rho_p_tilde} and \eqref{junction_cond_solved} in terms of $a_0$ and $c_0$ (and their time-derivatives), this constitutes a system of \textit{four} second order differential equations (with respect to $ t $) for the \textit{six} unknown functions $\mean{a'}, \mean{c'}, \mean{n'}, a_0, c_0$ and $\rho$. (The pressure functions $p$ and $p_{\phi}$ are not independent since they are linked to $\rho$ via an equation of state.) Consequently, the system is not closed and does not allow to determine the on-brane evolution in a deterministic way by simply fixing certain initial values at the position of the brane.

There is a very simple physical explanation for this failure: The evolution on the brane strongly depends on the wave content in the bulk. One can prepare gravitational waves (nonlinear wave solutions of the vacuum Einstein equations) in the bulk at some initial moment of time $t_i$. These waves will propagate towards the brane and eventually affect the on-brane evolution. In parlance of differential equations  this is equivalent to fixing certain initial conditions on a space-like hypersurface in the bulk, i.e. $h(t_i,y)=f_h(y)$ as well as $\dot h(t_i, y)=g_h(y)$ for $h \in \{a, c, n\}$, subject to the constraints \eqref{tt} and \eqref{ty}. The whole evolution in the bulk {\it and} on the brane is then uniquely determined by the system of PDEs \eqref{full_system}.  Of course, in general the on-brane evolution will depend on $f_h(y)$ and $g_h(y)$ with $y\neq 0$ and it is not sufficient to fix initial values only at the position of the brane.

However, there is one prominent exception. For codimension one, which is the DGP model, this approach allows to derive a closed on-brane system \cite{Deffayet:2000uy}. One might wonder why this is possible because also in this case the brane-evolution should be influenced by the absorption of bulk waves. The solution to this puzzle is quite simple: In Section \ref{sec:digression_codim_1} we confirm that in this particular setup the Einstein equations do not allow for wave solutions in the bulk that are compatible with the symmetries of the system. Thus, in this case the evolution is uniquely determined (up to one arbitrary constant of integration corresponding to a \textit{static} curved bulk geometry \cite{Deffayet:2000uy}) by fixing initial values solely at the position of the brane. This can also be understood as a consequence of a generalization of Birkhoff's theorem to geometries with planar symmetry \cite{Taub:1951}.

\section{Closing the on-brane system}
\label{sec:closed_system}

Now the question is whether it is still possible, by making some assumptions about the bulk geometry (but without solving for it), to arrive at a closed system of ODEs for the on-brane evolution. Since we are dealing with an empty bulk, a necessary condition we would like to impose is the absence of incoming gravitational waves from the bulk. As just argued, such incoming waves are the reason that the on-brane system is not closed. Therefore, we expect that imposing an ``outgoing wave condition'' will yield a closed system. This expectation is also consistent with the number of equations, because excluding incoming waves from the bulk on both sides of the brane would add two more equations to the system \eqref{ODE_full}, which is exactly the number that is needed to get a closed system.

Let us from now on simplify the analysis by assuming a reflection symmetry $ y \mapsto -y $ around the brane. For an empty bulk and the topology we are interested in, this assumption is quite natural. Regarding the on-brane system \eqref{ODE_full}, it amounts to setting all mean values to zero. Consequently, equations~\eqref{rr_ODE_J} and \eqref{tr_ODE_M} are satisfied identically. Recalling that \eqref{tr_ODE_J} is equivalent to the energy conservation equation \eqref{eq:en_cons} that allows to determine $ \rho $, only equation \eqref{rr_ODE_J} is left, which now reduces to
\begin{equation}
\label{eq:onBrane_symm}
	\frac{3}{4} \left[a'\right] \Bigl( \jump{a'} + \jump{c'} + \jump{n'} \Bigr) +\frac{1}{4} \jump{c'} \jump{n'} -3 \dot H_a-\dot H_c - 6 H_a^2-3 H_a H_c - H_c^2 = 0 \,.
\end{equation}
The remaining variables are $ a_0 $ and $ c_0 $, so the system is still under-determined, as expected. Due to the $ y $-symmetry, an outgoing wave condition would now only add one equation and should thus again yield a closed system.

The gravitational wave content of a metric is encoded in the Weyl tensor. In order to formulate an outgoing wave condition, it is therefore natural to ask whether the Weyl tensor can be decomposed in such a way that certain components can be identified as incoming or outgoing waves.
And indeed, it is possible --- at least in certain situations --- to find such a decomposition of the Weyl tensor, which can be obtained by looking at the geodesic deviation it causes for nearby freely falling test particles. This was first done in four spacetime dimensions by Szekeres \cite{Szekeres:1965ux} and more recently generalized to arbitrary dimensions by Podolsky and Svarc \cite{Podolsky:2012he}. The result is that the Weyl tensor can be decomposed into components corresponding to transverse and longitudinal, outgoing and incoming gravitational waves, as well as Newton-like parts.

Note that it was recently realized \cite{Hofmann:2013zea} that this interpretation is in fact not applicable to cylindrical geometries. More precisely, it was shown that in four-dimensional geometries with whole-cylinder symmetry, static configurations give rise to non-vanishing wave-like field components, and dynamic Einstein-Rosen waves can not be separated into incoming and outgoing parts by this procedure.
In our set-up, however, the expected gravitational waves are plane waves, for which the decomposition of the Weyl tensor was confirmed in \cite{Hofmann:2013zea} to work perfectly well.

\subsection{Interpretation of Weyl components}
\label{sec:interpret_weyl}

We will now briefly review the aforementioned decomposition of the Weyl tensor in $ d $ spacetime dimensions \cite{Szekeres:1965ux, Podolsky:2012he}, and then apply it to the concrete metric that we are interested in. In this section, capital Latin indices $ A, B, \ldots $ denote $ d $-dimensional spacetime indices and small Latin indices $ a, b, \ldots $ correspond to the $ d $ tetrad indices; $ d $-dimensional vectors are written in boldface.

First, one considers a time-like geodesic with unit tangent vector $ \mathbf{t} $ and chooses an orthonormal space-like vector $ \mathbf{x} $, which will correspond to the direction of wave-propagation. These two vectors are combined into two null vectors
\begin{align}
	\mathbf{k} := \frac{1}{\sqrt{2}} \left( \mathbf{t} + \mathbf{x} \right), && 	\mathbf{l} := \frac{1}{\sqrt{2}} \left( \mathbf{t} - \mathbf{x} \right),
\end{align}
and complemented with $ d-2 $ further orthonormal vectors $ \mathbf{m}_p $ $ (p = 2, \ldots, d-1) $ to form a real null tetrad, or  \textit{mixed tetrad}\footnote{In 4D, it is possible to construct a complete null tetrad by taking complex combinations of the space-like vectors \cite{Szekeres:1965ux}. But this is not necessary, and does not work for odd numbers of dimensions.}
\begin{equation}
\label{eq:def_mixed_tet}
	\mathbf{m}_a = (\mathbf{m}_0,\, \mathbf{m}_1,\, \mathbf{m}_p) := (\mathbf{k},\, \mathbf{l},\, \mathbf{m}_p).
\end{equation}
The indices $ \{ p, q, \ldots \} $ will always correspond to the $ d-2 $ space-like tetrad components and have the range $ (2, \dots, d-1) $.
By construction, these tetrad vectors now satisfy the quasi orthonormality relations
\begin{subequations}
	\begin{align}
		\mathbf{k}\cdot\mathbf{k} = \mathbf{l}\cdot\mathbf{l} &= 0,& \mathbf{k} \cdot \mathbf{l} &= -1 \\
		\mathbf{m}_p \cdot \mathbf{k} = \mathbf{m}_p \cdot \mathbf{l} &= 0, & \mathbf{m}_p\cdot\mathbf{m}_q &= \delta_{pq}.
	\end{align}
\end{subequations}
This frame is useful because certain components of the Weyl tensor\footnote{The Weyl tensor is defined as the traceless part of the Riemann tensor, i.e. $ C_{ABCD} := R_{ABCD} - \frac{2}{d-2} \left( R_{A[C} g_{B]D} - R_{B[C} g_{D]A} \right) + \frac{2}{(d-1)(d-2)} R\, g_{A[C}g_{D]B} $ in $ d $ spacetime dimensions.} in this frame,
\begin{equation}
	C_{abcd} = C_{ABCD}\, m^A_a m^B_b m^C_c m^D_d,
\end{equation}
can be given a physical interpretation by looking at the geodesic deviation they induce for nearby freely falling test particles. For instance, the term $ \Omega_{pq} := C_{0p0q} $ will deform a sphere of test particles lying in the hyperplane orthogonal to $ \mathbf{x} $ into an ellipsoid in this hyperplane and is therefore interpreted as a transverse gravitational wave component. The result is summarized in Table~\ref{tab:interpret_weyl}. However, note that this interpretation fails in certain cases \cite{Hofmann:2013zea}, as already mentioned. The components
\begin{equation}
\label{eq:weyl_non_obs}
	C_{0pqr}, C_{pqrs}, C_{01pq}  \; \text{and} \; C_{1pqr},
\end{equation}
which do not appear in this list, have no effect on the geodesic deviation at linear order and are thus not observable at this level.

\begin{table}[ht]
\centering
\begin{tabular}{ | l | l | l | l | }
\hline
\textbf{Component} & \textbf{Name} & \textbf{Identities} & \textbf{Interpretation} \\
\hline \hline
\multirow{2}{*}{$ C_{0p0q} $} & \multirow{2}{*}{$ \Omega_{pq} $} & $ \Omega_{pq} = \Omega_{qp} $ &
Transverse gravitational wave propagating \\
& & $ {\Omega^p}_p = 0 $ & in the direction  $ -\mathbf{x} $ \\
\hline
\multirow{2}{*}{$ C_{010p} $} & \multirow{2}{*}{$ \Psi_p $} & & Longitudinal gravitational wave propagating\\
& & & in the direction $ -\mathbf{x} $ \\
\hline
$ C_{0101}$ & $ \Phi $ & \multirow{2}{*}{$ \Phi + {\Phi^p}_p = 0 $} & \multirow{2}{*}{Newton-like part of the gravitational field} \\
$ C_{0p1q} $ & $ \Phi_{pq} $ & & \\
\hline
\multirow{2}{*}{$ C_{101p} $} & \multirow{2}{*}{$ \Psi'_p $} & & Longitudinal gravitational wave propagating\\
& & & in the direction $ +\mathbf{x} $ \\
\hline
\multirow{2}{*}{$ C_{1p1q} $} & \multirow{2}{*}{$ \Omega'_{pq} $} & $ \Omega'_{pq} = \Omega'_{qp} $ & Transverse gravitational wave propagating\\
& & $ {\Omega'^p}_p = 0 $ & in the direction  $ +\mathbf{x} $ \\
\hline
\end{tabular}
\caption{\label{tab:interpret_weyl}
The standard interpretation \cite{Szekeres:1965ux, Podolsky:2012he} of the components of the Weyl tensor. The indices $ (0,1, p) $ correspond to the mixed tetrad indices as in equation \eqref{eq:def_mixed_tet}, and the notation for the individual components follows \cite{Durkee:2010xq}. The listed identities follow directly from the symmetries and tracelessness of the Weyl tensor. For cases where the given interpretation fails, see \cite{Hofmann:2013zea}.
}
\end{table}

\subsection{Digression to codimension one}
\label{sec:digression_codim_1}

Before we investigate the six dimensional scenario, let us first take a look at the 5D case, i.e.\ the DGP model. The five dimensional metric describing a cosmological evolution on the brane can be written in the form
\begin{equation}
\label{eq:metric_5D}
	\mathrm ds^2 = -\re^{2n(t,y)} \mathrm dt^2 + \re^{2a(t,y)} \delta_{ij}\mathrm dx^i \mathrm dx^j + \mathrm dy^2.
\end{equation}
One obvious choice of a mixed tetrad associated with this metric is\footnote{Note that from now on the index range for $ \{ i, j, \ldots \} $ should be understood to be $ (2,3,4) $ in order to be consistent with the index range for the tetrad components.}
\begin{equation}
\label{eq:mixed_tet_5D}
	\mathbf{k} = \frac{1}{\sqrt{2}}\left( \re^{-n} \boldsymbol{\partial}_t + \boldsymbol{\partial}_y\right),\quad
	\mathbf{l} = \frac{1}{\sqrt{2}}\left( \re^{-n} \boldsymbol{\partial}_t - \boldsymbol{\partial}_y\right),\quad 
	\mathbf{m}_i =  \re^{-a} \boldsymbol{\partial}_{i},
\end{equation}
where we chose $ \mathbf{x} = \boldsymbol{\partial}_y $, which is the spatial direction perpendicular to the brane. However, there is a problem with this choice of tetrad: For the interpretation discussed above to apply, the time-like vector must be tangent to a geodesic. But $ \re^{-n} \boldsymbol{\partial}_t $ is in general \textit{not} parallel transported along its integral curves, so the frame \eqref{eq:mixed_tet_5D} can actually not be used. However, since the metric \eqref{eq:metric_5D} admits the Killing vectors $ \mathbf{m}_i $, it is clear that there are geodesics with tangent vectors of the form $ \mathbf{t} = f \boldsymbol{\partial}_t + g \boldsymbol{\partial}_y $ with some functions $ f(t,y), g(t,y) $. The various orthonormality relations among the vectors then imply that the resulting null vectors will have the form $ \alpha^{\pm 1} \left[ \re^{-n} \boldsymbol{\partial}_t \pm \boldsymbol{\partial}_y \right] $ with some function $ \alpha(t,y) $. Therefore, the Weyl components in the frame \eqref{eq:mixed_tet_5D}, will only differ from the ones for which the physical interpretation was derived by some overall (nonzero) factors. But since the main purpose is to set some of those components equal to zero, we do not care about those factors and will thus use the frame \eqref{eq:mixed_tet_5D} in the following.

Having established the frame, it is now straightforward to compute the various Weyl components. It turns out that \textit{all of the wave components vanish identically}:
\begin{equation}
	\Omega_{ij} = \Psi_i = 0 = \Omega'_{ij} = \Psi'_i.
\end{equation}
This means that the symmetries we assumed (3D isotropy and homogeneity) do not admit any propagating waves in the case of one extra dimension\footnote{This can also be seen from the well known fact that there is a generalized version of Birkhoff's theorem to geometries with planar symmetry \cite{Taub:1951}.}. The only non-vanishing components are the Newton-like fields. Due to the symmetry and the traceless condition they are not all independent. In fact, there is only one independent component:

\begin{align}
\label{eq:coul_5D}
	\Phi = 3 \left( {a'}^2 - \re^{-2n} \dot{a}^2 \right), &&
	\Phi_{ij} = -\frac{1}{3} \Phi\, \delta_{ij}.
\end{align}
Note that this term was simplified by using the vacuum Einstein equations to eliminate all second $ y $-derivatives as well as the second $ t $-derivative of $ a $.

(For completeness, it should be mentioned that there are also some non-vanishing Weyl components in the non-observable sector \eqref{eq:weyl_non_obs}. They are given by
\begin{equation}
	C_{ijkl} = -\frac{1}{3} \Phi \left( \delta_{ik}\delta_{jl} - \delta_{jk}\delta_{il} \right),
\end{equation}
so they contain no further independent components. Their appearance follows from the Newton-like terms \eqref{eq:coul_5D} by the traceless condition $ {C^{i}}_{jik} = C_{0j1k} + C_{1j0k} = \Phi_{jk} + \Phi_{kj} $.)

The impossibility of gravitational waves in the bulk is an important feature, which is special to the codimension one case. It implies that one can find a closed system of ODEs governing the on-brane evolution, without making any further assumptions about the bulk geometry. If bulk waves were possible, one could always prepare initial conditions in the bulk which could propagate towards the brane, leading to any arbitrary on-brane evolution. This is in agreement with the results of \cite{Binetruy:1999hy, Deffayet:2000uy}, where it was shown that the bulk Einstein field equations, together with the brane matching conditions suffice to derive a unique 4D modified Friedmann equation, containing only one arbitrary constant $ \mathcal{C} $. In fact, our findings show why this program was possible at all. Furthermore, $ \mathcal{C} $ vanishes if and only if the Newton term $ \Phi $ (and thus the Weyl tensor) is zero, confirming the physical interpretation of the constant as the Schwarzschild mass parameter of the bulk geometry \cite{Deffayet:2000uy}. Indeed, taking the mean of $ \Phi = 0 $ and assuming $ \langle a' \rangle = 0 $ (due to symmetry), together with the junction condition immediately gives
\begin{equation}
\label{eq:mod_Fried_codim_1}
	H_a = - \frac{\epsilon}{6 M_5^3} \tilde \rho^{(5)} \equiv - \frac{\epsilon}{6 M_5^3} \left( \rho - 3 M_4^2 H_a^2 \right),
\end{equation}
which is exactly the modified Friedmann equation of \cite{Deffayet:2000uy} with $ \mathcal{C} = 0 $ (and $ k = \Lambda = 0 $), and $ \epsilon = \pm 1 $ chooses the branch of the solution.

\subsection{Excluding incoming waves}
\label{sec:noInWaves}

Let us now come back to the case of two extra dimensions, where the metric in Gaussian normal coordinates reads
\begin{equation}
\label{eq:metric_6D_c}
	\mathrm ds^2 = -\re^{2n(t,y)} \mathrm dt^2 + \re^{2a(t,y)} \delta_{ij}\mathrm dx^i \mathrm dx^j + \mathrm dy^2 + \re^{2c(t,y)} \mathrm d\phi^2.
\end{equation}
The mixed orthonormal tetrad for this metric is
\begin{equation}
	\mathbf{k} = \frac{1}{\sqrt{2}}\left(\re^{-n} \boldsymbol{\partial}_t + \boldsymbol{\partial}_y\right),\quad
	\mathbf{l} = \frac{1}{\sqrt{2}}\left(\re^{-n} \boldsymbol{\partial}_t - \boldsymbol{\partial}_y\right),\quad 
	\mathbf{m}_i = \re^{-a} \boldsymbol{\partial}_{i}, \quad \mathbf{m_\phi} = \re^{-c} \boldsymbol{\partial}_\phi.
\end{equation}
Again, the physical interpretation of section \ref{sec:interpret_weyl} does not directly apply to this frame, because $ \re^{-n} \boldsymbol{\partial}_t $ is not tangent to a geodesic. But repeating the reasoning of section~\ref{sec:digression_codim_1}, one finds that this will only change the corresponding Weyl components by overall factors, which we are not interested in anyway. The straightforward calculation then gives:

\begin{subequations}
\begin{align}
	\Omega_{ij} &= \Oin \delta_{ij}, &\Omega_{\phi\phi} &= -3\Oin,	&\Omega_{i\phi} &= 0\\
	\Psi_{i} &= 0\\
	\Phi_{ij} &= \Phi^{(x)} \delta_{ij}, &\Phi_{\phi\phi} &= \Phi^{(\phi)} &\Phi &= -3\Phi^{(x)} - \Phi^{(\phi)} \\
	\Psi'_i &= 0\\
	\Omega'_{ij} &= \Oout \delta_{ij}, &\Omega'_{\phi\phi} &= -3\Oout,	&\Omega'_{i\phi} &= 0
\end{align}
\end{subequations}
with
\begin{subequations}
\label{eq:weyl_comp_6D}
\begin{align}
		\Oin &= 
		-\frac{a' c'}{2}-\frac{c' n'}{3} -\frac{\re^{-n}}{4} \left( a' \dot{a}-c' \dot{c} -n'\dot{a}+n'\dot{c} +\dot{a}'-\dot{c}' \right) + \frac{\re^{-2n}}{6} \left( 3\dot{a}\dot{c}+2 \dot{c}^2-2 \dot{c}\dot{n} + 2 \ddot{c}\right)
	\\
	\label{eq:coul_x}
		\Phi^{(x)} &= 
		- {a'}^2 -\frac{a' c'}{2} + \frac{\re^{-2 n}}{2} \left(2\dot{a}^2+ \dot{a} \dot{c}\right)
	\\
	\label{eq:coul_phi}
		\Phi^{(\phi)} &= \frac{3}{2} \left(- a' c' + \re^{-2 n} \dot{a} \dot{c} \right)
	\\
		\Oout &= 
		-\frac{a' c'}{2}-\frac{c' n'}{3} +\frac{\re^{-n}}{4} \left( a' \dot{a}-c' \dot{c} -n'\dot{a}+n'\dot{c} +\dot{a}'-\dot{c}' \right) + \frac{\re^{-2n}}{6} \left( 3\dot{a}\dot{c}+2 \dot{c}^2-2 \dot{c}\dot{n} + 2 \ddot{c}\right)
\end{align}
\end{subequations}
which were again simplified by using the vacuum Einstein equations to eliminate all second $ y $-derivatives as well as the second $ t $-derivative of $ a $.

(As in 5D, there are also some non-zero Weyl components in the class of terms \eqref{eq:weyl_non_obs}. But again, they do not contain any new independent terms, and their appearance is required by the traceless conditions of the Weyl tensor.)

So in a six dimensional geometry with (spatially flat) 3D isotropy and homogeneity on the brane as well as rotational symmetry around the compact extra dimension, the Weyl tensor is completely characterized by the four components \eqref{eq:weyl_comp_6D}. According to the standard interpretation, they correspond to left-moving (i.e.\ propagating in direction $ -\mathbf{y} $) gravitational waves $ \Oin $, to two independent Newton-like field components $ \Phi^{(x)}, \Phi^{(\phi)} $, and to right-moving gravitational waves $ \Oout $. As an aside, note that the appearance of wave-terms is only possible because the $ \phi $-direction is different from the $ x $-directions in \eqref{eq:metric_6D_c}, reducing the symmetries of the metric. If one had set $ c = a $ in this ansatz, the wave parts would have vanished identically, as in the 5D case.

As already mentioned, we expect the interpretation of the Weyl components to be correct in our set-up. Therefore, to exclude incoming bulk waves, we set $ \Oin(y>0) = 0 $ and $ \Oout(y<0) = 0 $. To convert these conditions into on-brane equations, we simply take the limit $ y \to 0^+ $ and $ y \to 0^- $, respectively. Due to the $ Z_2 $ symmetry, this yields one further on-brane equation:
\begin{multline}
\label{eq:onBrane_noInWaves}
	\jump{a'}\jump{c'} + \frac{2}{3}\jump{c'}\jump{n'} + H_a \left( \jump{a'} - \jump{n'} \right) - H_c \left( \jump{c'} - \jump{n'} \right) + \dot{\jump{a'}} - \dot{\jump{c'}} \\
	- \frac{4}{3} \left( 2\dot H_c + 3 H_a H_c + 2H_c^2 \right) = 0
\end{multline}
After using \eqref{junction_cond_solved} to eliminate all the jumps, the two equations \eqref{eq:onBrane_symm} and \eqref{eq:onBrane_noInWaves} indeed constitute a closed on-brane system of ODEs that in principle allows to solve for $ a_0(t) $ and $ c_0(t) $. Note that it is second order in both $ a_0 $ and $ c_0 $, and so one has to specify four initial conditions. However, we expect at least one of these constants (or some combination of them) to correspond to a parameter measuring the constant curvature of the bulk geometry. This also happens in the DGP case, with the (technical) difference that there the second order ODE can be integrated once analytically, in which case the corresponding constant of integration turns out to be the Schwarzschild mass parameter $ \mathcal{C} $ of the bulk geometry \cite{Deffayet:2000uy}. The subsequent analysis of cosmological solutions is then usually simplified by setting $ \mathcal{C} = 0 $.

In the case at hand, an analytic integration of the ODEs does not seem feasible. However, the decomposition of the Weyl tensor still allows to identify the Newton-like parts of the bulk gravitational field, viz.\ \eqref{eq:coul_x} and  \eqref{eq:coul_phi}. Therefore, setting them to zero should be analogous to setting $ \mathcal{C} = 0 $ in the DGP case. One might worry that this could add an additional independent on-brane equation thus leading to an over-determined system. However, as we will show now, this is not the case.

\subsection{Zero Newton-like bulk fields}

We will now assume that the bulk gravitational field does not contain any Newton-like components,
\begin{align}
\label{eq:coul_0}
	\Phi^{(x)} = 0, \quad \Phi^{(\phi)} = 0 && \left( y \neq 0 \right).
\end{align}
Due to the mirror symmetry\footnote{Without assuming this symmetry, it turns out that the jumps of \eqref{eq:coul_0} together with \eqref{rr_ODE_J} imply --- if $ \jump{a'}, \jump{c'} \neq 0 $, which would otherwise yield $ \tilde\rho = 0 $ --- that all the means have to vanish. In other words, zero Newton-like field components can only be achieved for (locally) symmetric configurations.} in $ y $, these conditions add two on-brane equations, which can be brought into the following form:
\begin{subequations}
\label{eq:mod_Fried_coul_0_jumps}
	\begin{align}
	\label{eq:mod_Fried_a_coul_0}
		2 H_a & = \sigma \jump{a'} \,, \\
	\label{eq:mod_Fried_c_coul_0}
		2 H_c & = \sigma \jump{c'} \,,
	\end{align}
\end{subequations}
where $ \sigma = \pm 1 $. (Note that in deriving \eqref{eq:mod_Fried_c_coul_0}, we divided by $ H_a $, so for a static solution this equation would be absent.)  Without making any assumptions about the wave components, for the moment, there are now three on-brane equations: \eqref{eq:onBrane_symm}, \eqref{eq:mod_Fried_a_coul_0} and \eqref{eq:mod_Fried_c_coul_0}. However, it turns out that after using \eqref{eq:mod_Fried_coul_0_jumps} in \eqref{eq:onBrane_symm}, it becomes identical to \eqref{tr_ODE_J} and thus to the energy conservation equation \eqref{eq:en_cons}. \textit{As a result, the two equations \eqref{eq:mod_Fried_coul_0_jumps} constitute a consistent, closed system of ODEs for $ a_0(t) $ and $ c_0(t) $.}

This is rather surprising, because so far we have not made any assumptions about the wave content in the bulk, so a priori incoming waves could still be possible. Nonetheless, we arrived at a closed on-brane system. How can this be possible?
To answer this question, let us evaluate the wave components of the Weyl tensor at the brane position, using the relations~\eqref{eq:mod_Fried_coul_0_jumps} (as well as \eqref{tr_ODE_J}):
\begin{subequations}
\label{eq:wave_comp_coul_0}
\begin{gather}
\begin{align}
	\Oout(0^+) &= \Oin(0^-) = (1 - \sigma) \Omega(t) \\
	\Oout(0^-) &= \Oin(0^+) = (1 + \sigma) \Omega(t)
\end{align} \\
	\Omega(t) := \frac{H_a \left( \dot H_c + H_c^2 \right) - H_c \left( \dot H_a + H_a^2 \right)}{3 H_a+H_c}
\end{gather}
\end{subequations}
This shows that in the branch $ \sigma = -1 $ there are only waves propagating away from the brane, while the branch $ \sigma = +1 $ only allows for waves traveling towards the brane. Therefore, the absence of Newton-like field components implies that there are either purely incoming or purely outgoing waves, which allowed to arrive at the closed system \eqref{eq:mod_Fried_coul_0_jumps}.

We choose the $ \sigma = -1 $ branch, because we do not want any incoming bulk waves. After using \eqref{junction_cond_solved}, the modified Friedmann equations for this branch become
\begin{subequations}
\label{eq:mod_Fried_coul_0_tildes}
	\begin{align}
		H_a &= \frac{1}{8 M_6^4} \left( \tilde\rho + \tilde p - \tilde p_{\phi} \right) \,, \\
		H_c &= \frac{1}{8 M_6^4} \left( \tilde\rho - 3 \tilde p + 3 \tilde p_{\phi} \right) \,.
	\end{align}
\end{subequations}
Since the outgoing wave components vanish, all solutions of these equations are contained in the larger class of solutions to the system discussed in Section \ref{sec:noInWaves}. But they correspond only to those solutions for which the initial conditions are compatible with vanishing Newton-like field components. In that sense they are analogous to the $ \mathcal{C} = 0 $ solutions in the DGP model. However, while $ \mathcal{C} = 0 $ in the DGP case implies a completely flat bulk spacetime, here the situation is different because the geometry still allows for outgoing gravitational waves.

As already discussed, without setting the Newton-like components to zero the system of ODEs is second order for both $ a_0 $ and $ c_0 $. Now we see that --- without the brane induced gravity terms --- the system \eqref{eq:mod_Fried_coul_0_tildes} is first order for both variables, so only two initial conditions need to be specified, and the ``$ \text{Newton} = 0 $'' solutions correspond to a two-dimensional subspace of the four dimensional parameter space of initial conditions characterizing the most general outgoing wave solutions. The brane induced gravity terms, however, reintroduce second $ t $-derivatives. But they only enter via the combination $ \left( \dot H_a - \dot H_c \right) $, so there is one constraint left, and the number of required initial values equals three. Explicitly, the two modified Friedmann equations can for instance be written as:

\begin{subequations}
\begin{empheq}[box=\fbox]{align}
	3 H_a + H_c &= \frac{1}{2 M_6^4} \left[ \rho - 3M_5^3 \left( H_a^2 + H_a H_c \right)\right] \,,   \label{eq:mod_Fried1_coul_0} \\
	H_c &= \frac{1}{8 M_6^4} \left[ \rho - 3 p + 3 p_{\phi} + 3M_5^3 \left( \dot H_a - \dot H_c + 2H_a^2 - 3H_a H_c - H_c^2 \right) \right]  \label{eq:mod_Fried2_coul_0}
\end{empheq}	
\label{eq:mod_Fried_coul_0}
\end{subequations}

As will be shown in Appendix \ref{ap:full_bulk_sol}, the assumption \eqref{eq:coul_0} even allows to solve for the whole bulk geometry, confirming the picture of gravitational waves propagating off the brane into one infinite extra dimension, thus justifying the Weyl component procedure.

\subsection{Distinction from codimension two set-up}
\label{sec:distinction_codim_2}

The ansatz \eqref{eq:metric_6D_c} is formally the same that would be used for a radially symmetric spacetime, with $ y $ taking the role of a radial coordinate, cf.\ Figure \ref{fig:capped_cone}. It could therefore be used for a model in which the brane is a codimension two object living in two infinite extra dimensions. If the brane is regularized by blowing it up to a circle of finite radius, then also the action would formally take the same form as here, equation \eqref{ActionBIG}.

The solution that we derived, however, does not correspond to such a geometry, but rather to the cylindrical geometry depicted in Figure~\ref{fig:cylinder}. This can be seen in different ways: First of all, there is in general no axis --- a place where $ c=0 $ --- in our spacetime, as would be necessary for a cylindrical geometry. Furthermore, even if there were an axis, say, located at some $ y_{\rm a} < 0 $, then due to the mirror symmetry there would also be an axis at $ -y_{\rm a} > 0 $. More generally, the mirror symmetry is incompatible with a distinction between an interior and exterior region as would be required in the codimension two set-up. (Note that we did not need to enforce the mirror symmetry by hand, it already followed from the assumption of zero Newton-like Weyl components.) Finally, the full bulk solution that is presented in Appendix~\ref{ap:full_bulk_sol} explicitly shows that the spacetime describes plane waves, consistent with the one infinite (and one compact) extra dimension that we are modeling, but inconsistent with a codimension two picture in which there would have to be cylindrical waves.

One also has to keep in mind that while the interpretation of the Weyl components works for planar symmetry, it fails for cylindrical geometries \cite{Hofmann:2013zea}. Hence, the outgoing wave criterion, which was rightfully used here in Section \ref{sec:noInWaves} to derive a closed on-brane system, could not be used for the codimension two case. In fact, one cannot expect to be able to derive a closed on-brane system in that case, because there is no local outgoing wave criterion for cylindrical waves, already at the linear level. Therefore, in the codimension two system one has to deal with PDEs (i.e.\ the full set of Einstein equations) instead of simple ODEs, which will be much more difficult to solve. The corresponding analysis is presented elsewhere~\cite{Niedermann:2014bqa}.

\section{Solutions}
\label{sec:solution}

\subsection{Stabilizing the compact extra dimension}

As already mentioned, the on-brane equations need to be supplemented with equations of state for both $ p $ and $ p_\phi $. For the former we shall assume a linear equation of state of the form $ p = w \rho $, with some constant $ w $. The latter, on the other hand, could for instance be used to stabilize the compact extra dimension. Let us now investigate the consequences of this scenario.

Stabilization means that we set $ H_c = 0 $. The azimuthal pressure $ p_\phi $ that is needed to achieve this, can be read off from equation \eqref{eq:mod_Fried2_coul_0} and allows to infer the corresponding equation of state 
\begin{align}\label{eq:DGPlike}
w_{\phi}:=\frac{p_{\phi}}{\rho}=\frac{-1+3w}{2}-\frac{1+w}{2\sqrt{1+\chi}}+\frac{4}{3\chi}\left(\sqrt{1+\chi}-1 \right)\;,
\end{align}
where $\chi:=M_5^3 \rho / (3 M_6^8)$. 
For a pure 4D cosmological constant source ($w=-1$) this equation together with $\chi \geq0$ implies $w_{\phi}\leq -\frac{4}{3}$. This shows that a stabilization requires unphysical matter leading to a violation of the Null Energy Condition. For dust and radiation we find $w_{\phi}\geq -1$ and $w_{\phi}\geq -\frac{2}{3}$, respectively. Therefore, it is possible to implement a stabilization in these two cases by means of physical matter.  In other words, the assumption of having a stabilized angular direction is only valid during a dust or radiation dominated epoch.

The corresponding evolution of $ a_0(t) $ is determined by \eqref{eq:mod_Fried1_coul_0}, which now takes the form
\begin{equation}\label{eq:mod_Fried_coul_0_c_const}
	3 H_a  = \frac{1}{2 M_6^4} \left( \rho - 3M_5^3 H_a^2 \right) \,.
\end{equation}
It turns out that \eqref{eq:mod_Fried_coul_0_c_const} is exactly the same modified Friedmann equation as in the DGP case~\eqref{eq:mod_Fried_codim_1}, after dimensionally reducing the energy density and the Planck masses according to
\begin{align}
\rho \to \frac{\rho^{(4)}}{2\pi R}, && M_5^3 \to \frac{M_4^2}{2\pi R}, && M_6^4 \to \frac{M_5^3}{2\pi R} \,,
\end{align}
where $ R := \exp(c_0) $, so that $ 2\pi R $ is the physical circumference of the compact extra dimension.

To be more precise, it is the \textit{normal branch} of the DGP cosmology. The \textit{self-accelerated branch} would be recovered by taking the $ c_0 = \text{constant} $ limit of the ``incoming wave'' solution ($ \sigma = -1 $). However, \eqref{eq:wave_comp_coul_0} immediately shows that the assumption $ H_c = 0 $ also implies that the wave-components  vanish, and the solution is in fact Riemann-flat\footnote{This could serve as a physically justified criterion for dismissing the self-accelerated branch of the DGP cosmology, if one is willing to view this branch as a limiting case of one additional, approximately constant, compact extra dimension, because in that case it would correspond to incoming bulk waves incompatible with a source-free bulk.}. But then the result that we reproduced the DGP equation can physically be understood because the only difference to the DGP case is the addition of one trivial compact extra dimension to the geometry, which does not take part in the dynamics of the model at all. 

Phenomenologically, these solutions are promising because for early times ($H_a r_c \gg 1$) they reproduce the standard Friedmann evolution in analogy with the DGP model. However, for late times ($H_a r_c \lesssim 1$) there are two sources of modification. First, the bulk term, i.e.\ the left side of \eqref{eq:mod_Fried_coul_0_c_const}, starts to dominate. This effect alone would lead to a modification which is not different from the one in the DGP model. But there is a second modification due to the fact that the compact dimension cannot be stabilized once the evolution enters the tension (i.e.\ cosmological constant) dominated epoch. The dynamics of the corresponding size modulus, described by $H_c$, as well as the presence of emitted bulk waves will affect the evolution significantly. Of course, in order to investigate this effect, it is necessary to resolve the stabilization mechanism dynamically. This could be done along the lines of \cite{Kaloper:2007ap} where an axion-like field, living on the brane, winds around the compact dimension in order to stabilize it. This approach is beyond the scope of the present work and requires further study.

There are of course other possible solutions in the class of \eqref{eq:mod_Fried_coul_0}, for which the size of the compact extra dimension is not stabilized at all, and which imply a different evolution of $ a_0 $. These solutions generically contain waves that are emitted into the bulk. As a specific representative of this class of solutions, we will consider the case of vanishing azimuthal pressure. 

\subsection{Zero azimuthal pressure}

In the case of vanishing azimuthal pressure, $ p_\phi = 0 $, the energy conservation \eqref{eq:en_cons} implies that $ \rho $ scales like
\begin{equation}
	\rho \propto \re^{-3 (1+w) a_0 - c_0} \,.
\end{equation}
Hence, the dimensionally reduced 4D energy density scales like
\begin{equation}
	\rho^{(4)} \equiv 2\pi R \, \rho \propto \re^{c_0} \rho \propto \re^{-3 (1+w) a_0} \,,
\end{equation}
that is, exactly like in standard 4D GR without any extra dimensions. In particular, choosing $ w=-1 $ then implies that $ \rho^{(4)} $ is constant, in accordance with the interpretation of a 4D cosmological constant. This makes the choice $ p_\phi = 0 $ very special and can be viewed as a motivation for studying it. In fact, it is the only choice --- apart from the stabilized scenario $ H_c = 0 $ --- which achieves this.

Denoting initial values by a subscript $ \rm{i} $, we need to specify $ \rho_{\rm{i}} $, $ a_{0\rm{i}} $, $ c_{0\rm{i}} $ and e.g.\ $ H_{a\rm{i}} $, in which case $ H_{c\rm{i}} $ is determined by the constraint \eqref{eq:mod_Fried1_coul_0}. We can, however, without loss of generality\footnote{The general solutions are obtained by substituting $ a_0(t) \to a_0(t) + a_{0\rm{i}} $ and $ c_0(t) \to c_0(t) + c_{0\rm{i}} $.}, assume $ a_{0\rm{i}} = c_{0\rm{i}} = 0 $. 
The solutions for $ a_0 $ and $ c_0 $ are then uniquely determined by \eqref{eq:mod_Fried_coul_0} for any given choice of $ H_{a\rm{i}} $, $ \bar \rho := \rho_{\rm i} / M_6^4 $ and $ r_c := M_5^3 / M_6^4 $.

Since the size of the compact dimension is not stabilized, we do not expect to have a pure 4D regime for this solution. More precisely, in the regime where the brane induced term dominates, the corresponding size modulus $c_0(t)$ affects the gravitational dynamics in a non-trivial way. This can be seen directly from \eqref{eq:mod_Fried_coul_0}, where only for $H_c=0$ the equations would take the standard Friedmann form in the limit $M_6 \rightarrow 0$. 

Let us first focus on pure dust ($ w = 0 $) and pure cosmological constant ($ w=-1 $) solutions, before we discuss the general fluid containing both components and its phenomenological viability.

\subsubsection{Pure dust}

For $ p = p_\phi = 0 $, there exist exact solutions with $ a_0(t) = c_0(t) $, satisfying
\begin{equation}
\label{eq:mod_Fried_a=c}
	H_a = \frac{1}{8} \bar\rho \, \re^{-4 a_0} - \frac{3}{2} r_c \, H_a^2 \,.
\end{equation}
This is equivalent (apart from numerical factors) to the DGP modified Friedmann equation, sourced by radiation. The existence of this class of solutions can easily be understood because by setting $ a = c $, the metric ansatz \eqref{eq:ind_met} can equally well describe a $ (4+1) $-dimensional, spatially homogeneous and isotropic brane of codimension $ 1 $. Thus, it is the straightforward generalization of the usual 4/5D DGP model to a 5/6D ``DGP'' set-up. (In fact, this class of solutions could be extended to $ p = p_\phi \neq 0 $, with $ \rho \propto \re^{-4 (1+w) a_0} $.) Furthermore, in this case the wave components \eqref{eq:wave_comp_coul_0} vanish and the brane is thus embedded in a flat bulk geometry.

\begin{figure*}[htb]
	\centering
	\begin{subfigure}{0.48\textwidth}
		\includegraphics[width=\textwidth]{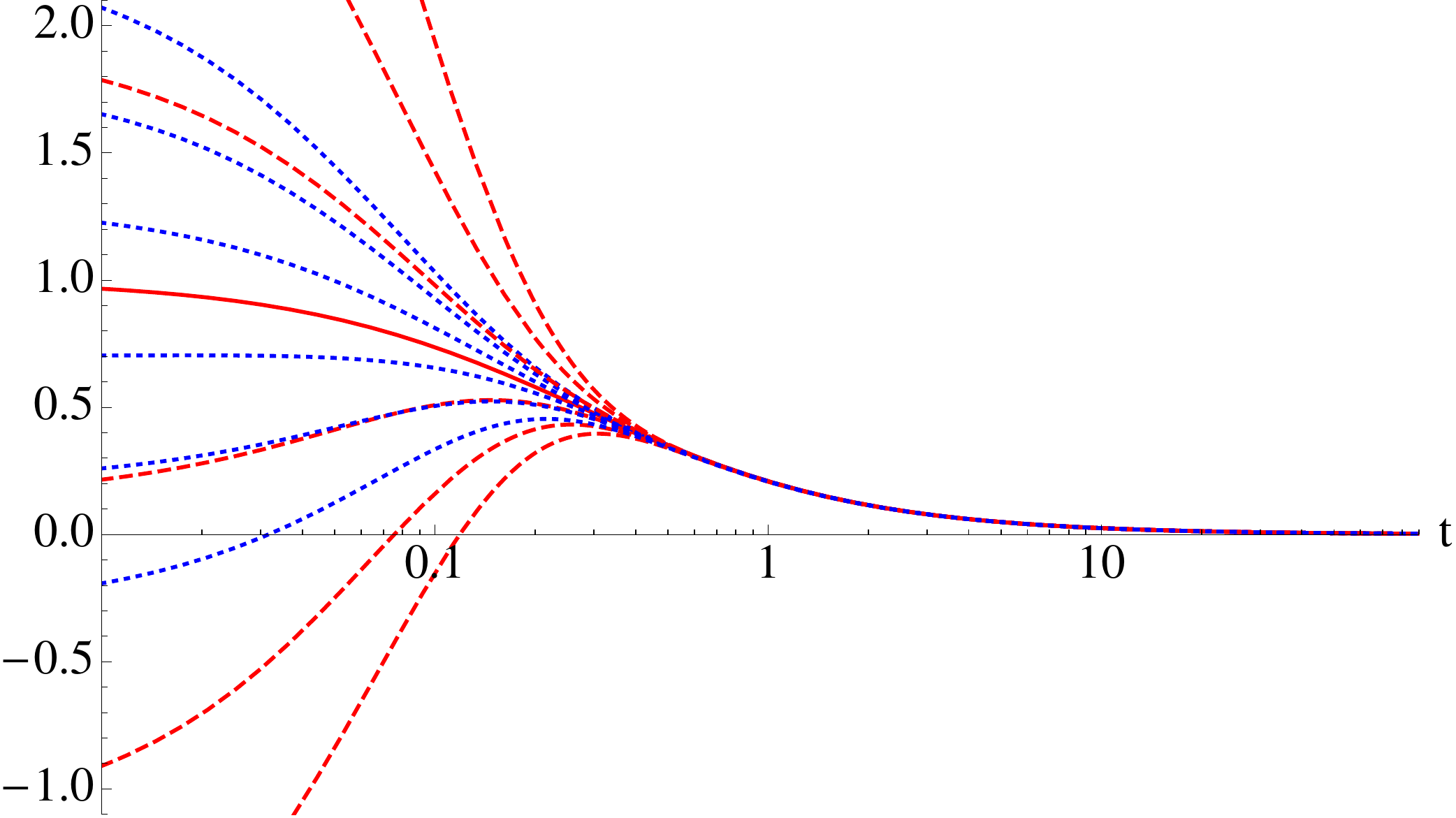}
		\caption{Dust}
		\label{fig:dust_attractor}
	\end{subfigure}
	\hfill
	\begin{subfigure}{0.48\textwidth}
		\includegraphics[width=\textwidth]{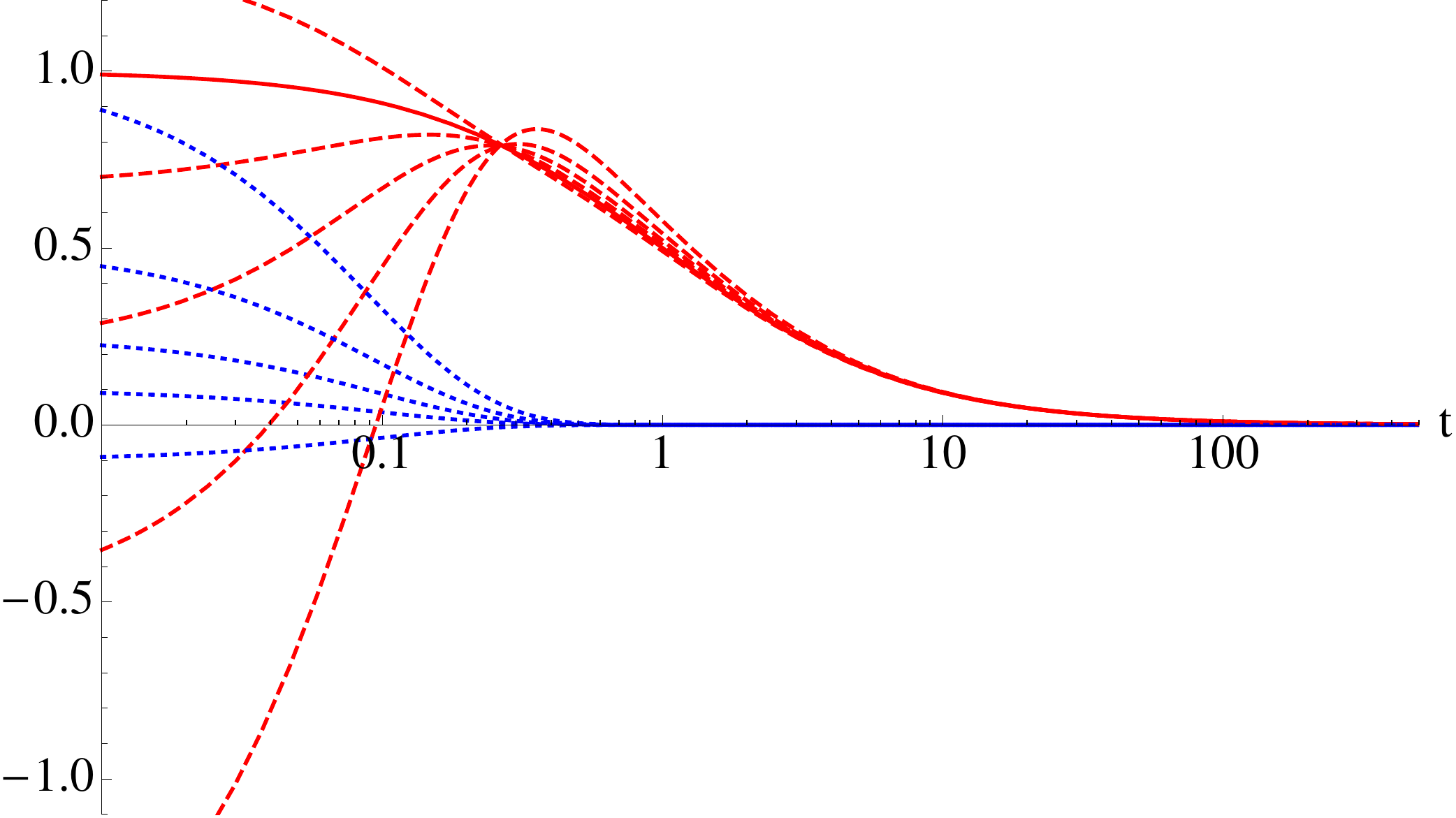}
		\caption{Cosmological constant}
		\label{fig:CC_attractor}
	\end{subfigure}%
	\caption{The Hubble parameters $ H_a $ (blue, dotted) and $ H_c $ (red, dashed) for different initial conditions generically approach the attractor solutions (solid lines). Time and the Hubble parameters are measured in the same fixed but arbitrary units.}
		\label{fig:attractors}
\end{figure*}

In this class of solutions, the initial conditions satisfy $ H_{a\rm{i}} = H_{c\rm{i}} $. For $ H_{a\rm{i}} \neq H_{c\rm{i}} $ the evolution will differ from the exact $ a_0 = c_0 $ one. But it turns out that it is an attractor solution, i.e.\ for late times the $ a_0 = c_0 $ solution is approached for generic initial conditions. This is shown in Figure~\ref{fig:dust_attractor}, where  some numerical solutions for $ H_a $ (blue, dotted) and $ H_c $ (red, dashed) are plotted for fixed $ r_c $ and $ \bar \rho $ but different $ H_{a\rm{i}} $ (and thus different $ H_{c\rm{i}} $, according to the constraint). The solid line corresponds to $ H_{a\rm{i}} = H_{c\rm{i}} $, and therefore $ a_0 = c_0 $ for all times. Note that while no gravitational waves are emitted into the bulk for the attractor solution, there are non-zero wave components (but approaching zero as $ t \to \infty $) for all other solutions.

\subsubsection{Pure cosmological constant}

For $ p = -\rho $ and $ p_\phi = 0 $, there are exact solutions of \eqref{eq:mod_Fried_coul_0} with
\begin{align}
\label{eq:degrav_sol}
	a_0 = 0 && \text{and} && c_0 = \log( 1 + \frac{\bar\rho}{2} t ) \,.
\end{align}
So in this case the expansion is solely in $ \phi $-direction. From a 4D point of view, this can be viewed as a realization of the so-called \textit{degravitation mechanism} \cite{Dvali:2002pe, Dvali:2002fz, ArkaniHamed:2002fu, Dvali:2007kt, deRham:2007rw}, because the Hubble parameter in $ \vec{x} $-direction is zero, despite the presence of a 4D cosmological constant $ \rho^{(4)} $. 

Again, one can verify that these are attractor solutions, as can be seen from Figure~\ref{fig:CC_attractor}. The dotted (blue) and dashed (red) lines show numerical results for $ H_a $ and $ H_c $, respectively, again for fixed $ r_c $ and $ \bar\rho $ but different $ H_{a\rm{i}} $ and $ H_{c\rm{i}} $. They asymptotically approach the solid lines, corresponding to the exact solution \eqref{eq:degrav_sol}. As with the dust solutions, the outgoing wave components vanish for the attractor solutions, but are non-zero for the other solutions.

\subsubsection{Dust and cosmological constant}

For a fluid containing both dust and a cosmological constant, the solutions can in general only be obtained numerically. Since the dust contribution to $ \rho $ falls off faster than that due to the cosmological constant, the late time asymptotic is that of the degravitating attractor~\eqref{eq:degrav_sol}. Depending on the other parameters, there can also be an intermediate regime in which the solution behaves like the $ a_0 = c_0 $ attractor.

\begin{figure}[hbt]
	\centering
	\begin{subfigure}{0.48\textwidth}
		\includegraphics[width=\textwidth]{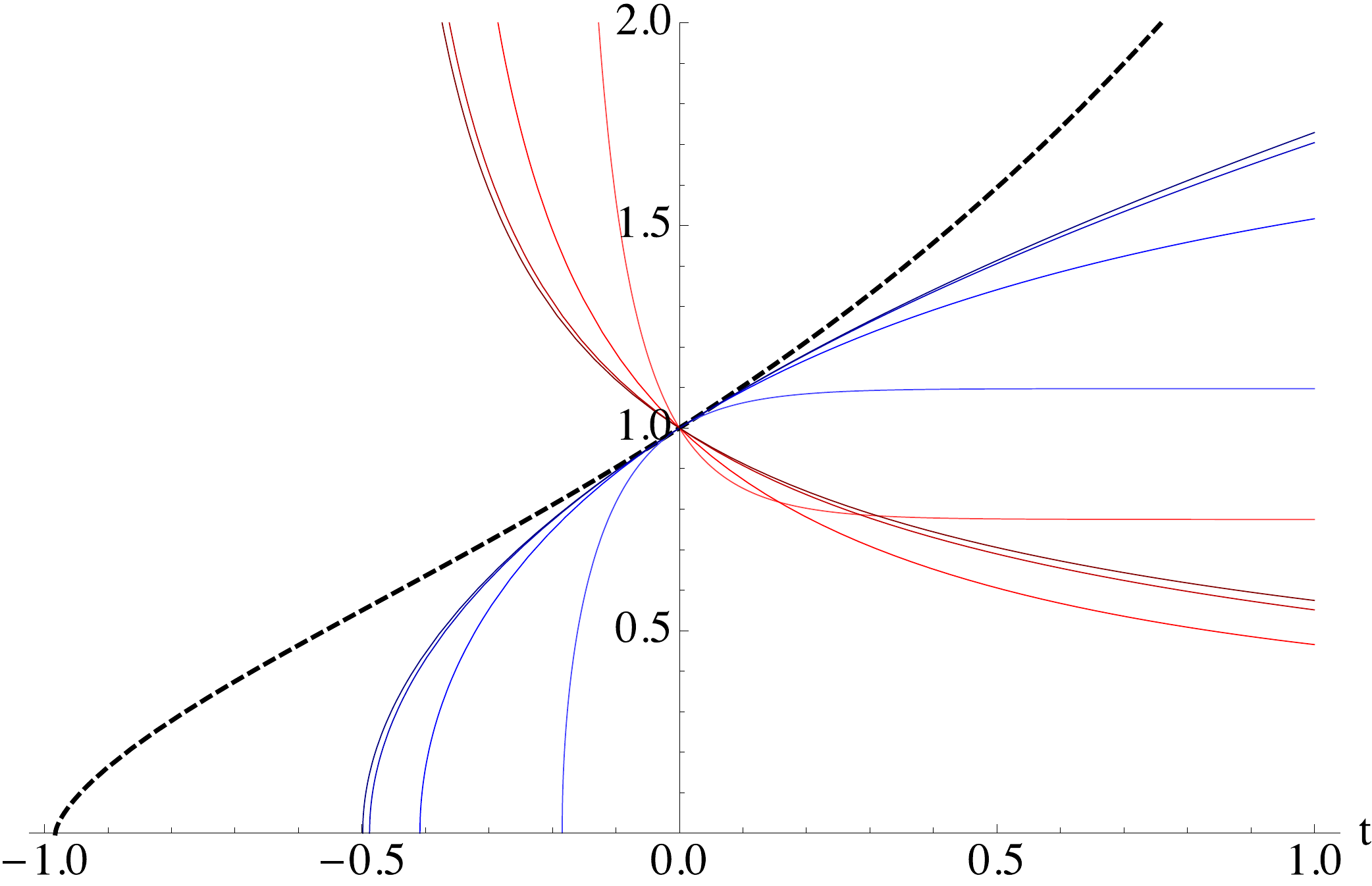}
		\caption{$ r_c \in \{0.1, 1, 10, 100 \} $, $ \bar\rho_{\rm{dust}} = \bar\rho_{\rm{cc}} = 0 $}
		\label{fig:vary_rc}
	\end{subfigure}\\
	\begin{subfigure}{0.48\textwidth}
		\includegraphics[width=\textwidth]{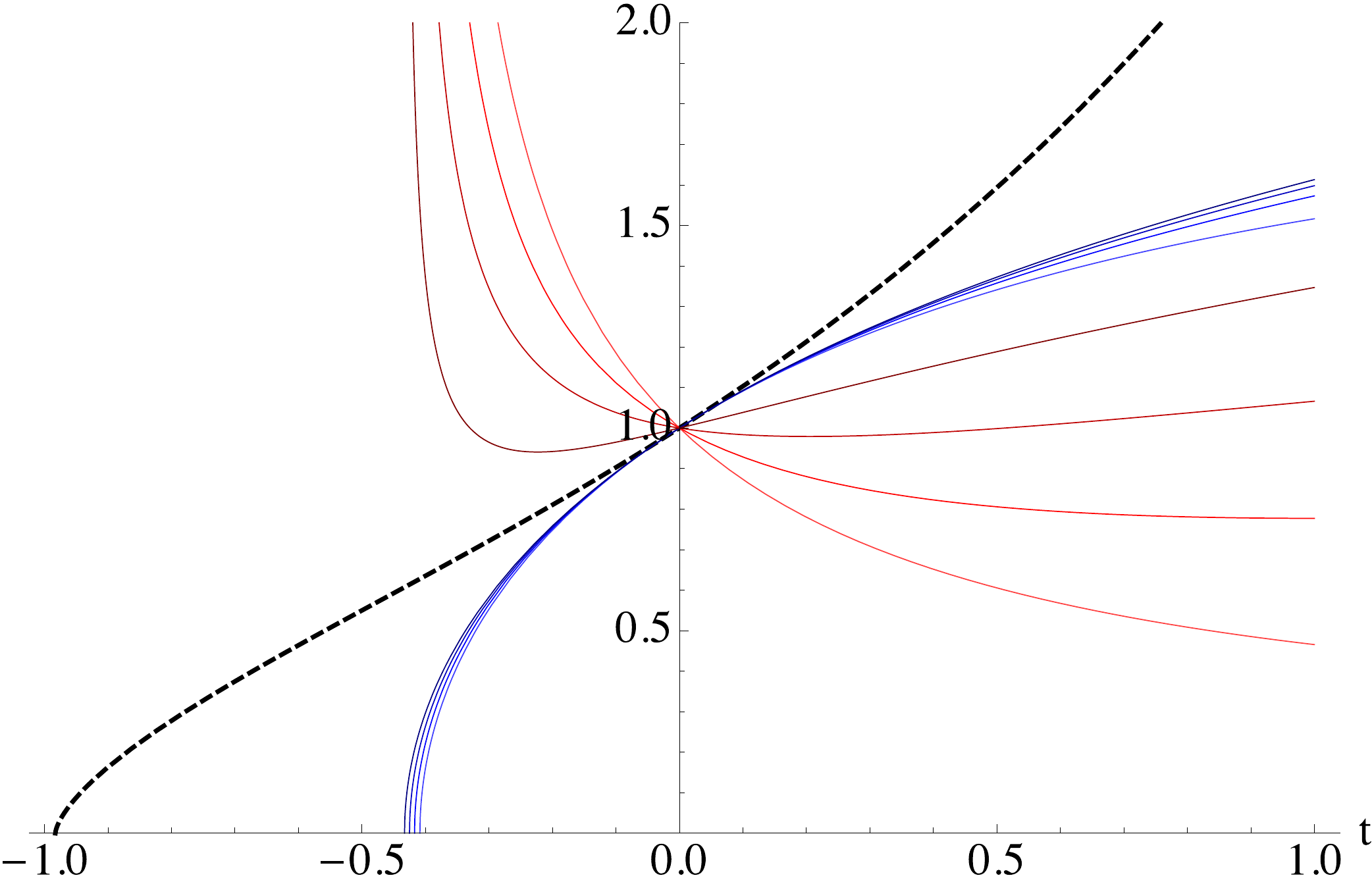}
		\caption{$ r_c = 1$, $ \bar\rho_{\rm{dust}} \in \{0, 5, 10, 15 \} $, $ \bar\rho_{\rm{cc}} = 0 $}
		\label{fig:vary_Omat}
	\end{subfigure}
	\hfill
	\begin{subfigure}{0.48\textwidth}
		\includegraphics[width=\textwidth]{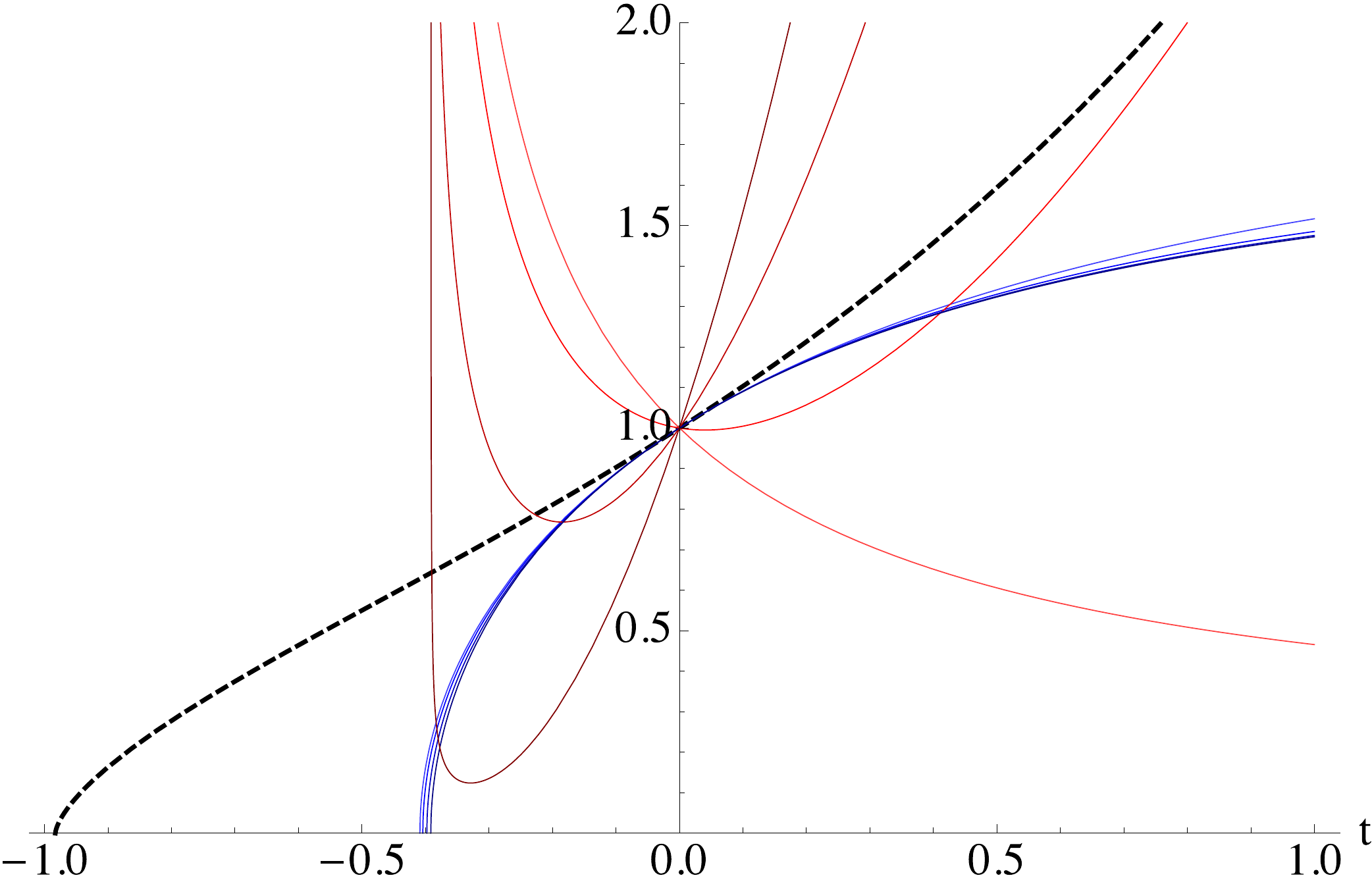}
		\caption{$ r_c = 1$, $ \bar\rho_{\rm{dust}} = 0 $, $ \bar\rho_{\rm{cc}} \in \{0, 10, 30, 50 \} $}
		\label{fig:vary_Occ}
	\end{subfigure}%
	\caption{The scale factors $ \re^{a_0} $ (blue) and $ \re^{c_0} $ (red) for various choices of parameters; the dashed black lines show the scale factor for a standard $ \Lambda $CDM cosmology with $ \Omega_\Lambda = 0.72 $. The lightest curves correspond to the first parameter in each list. All dimensionful quantities are measured in units of $ H_{a\rm{i}} $.}
		\label{fig:scaleFactors}
\end{figure}

These attractor solutions are clearly not capable of explaining the evolution history of the universe as we observe it. But it is possible to have solutions that are far away from the attractors for early times, which might in principle still be in accordance with cosmological observations. However, a numerical analysis shows that the past evolution is never close to the standard 4D $ \Lambda $CDM evolution. In Figure~\ref{fig:scaleFactors}, the scale factors $ \re^{a_0} $ (blue lines, $ \to 0 $ in the past) and $ \re^{c_0} $ (red lines, $ \to \infty $ in the past) are plotted for various parameters. The black dashed line is the scale factor for standard $ \Lambda $CDM with $ \Omega_{\Lambda} = 0.72 $. The crossover $ r_c $ has the greatest impact on the evolution of $ \re^{a_0} $, while $ \bar\rho_\mathrm{dust} $ and $ \bar\rho_\mathrm{cc} $ mainly influence $ \re^{c_0} $. 
Increasing $ \bar\rho_\mathrm{dust} $ and $ \bar\rho_\mathrm{cc} $ moves the blue curves closer towards the standard $ \Lambda $CDM curve, but if either parameter becomes too large, the angular scale factor $ \re^{c_0} $ becomes zero even before\footnote{We are looking backwards in time, so by ``before'' we mean larger $ t $ here.} the Big Bang in $ x $-direction is reached. Figure~\ref{fig:scaleFactors} only shows cases in which $ \bar\rho_\mathrm{dust} $ and $ \bar\rho_\mathrm{cc} $ are small enough so that $ \re^{c_0} $ turns around before it reaches zero, leading to a Kasner-like behavior close to the singularity. Hence, it is not possible to get much closer to the standard evolution than shown in these plots by further increasing $ \bar\rho_\mathrm{dust} $ or $ \bar\rho_\mathrm{cc} $.

\begin{figure}
	\includegraphics[width=\textwidth]{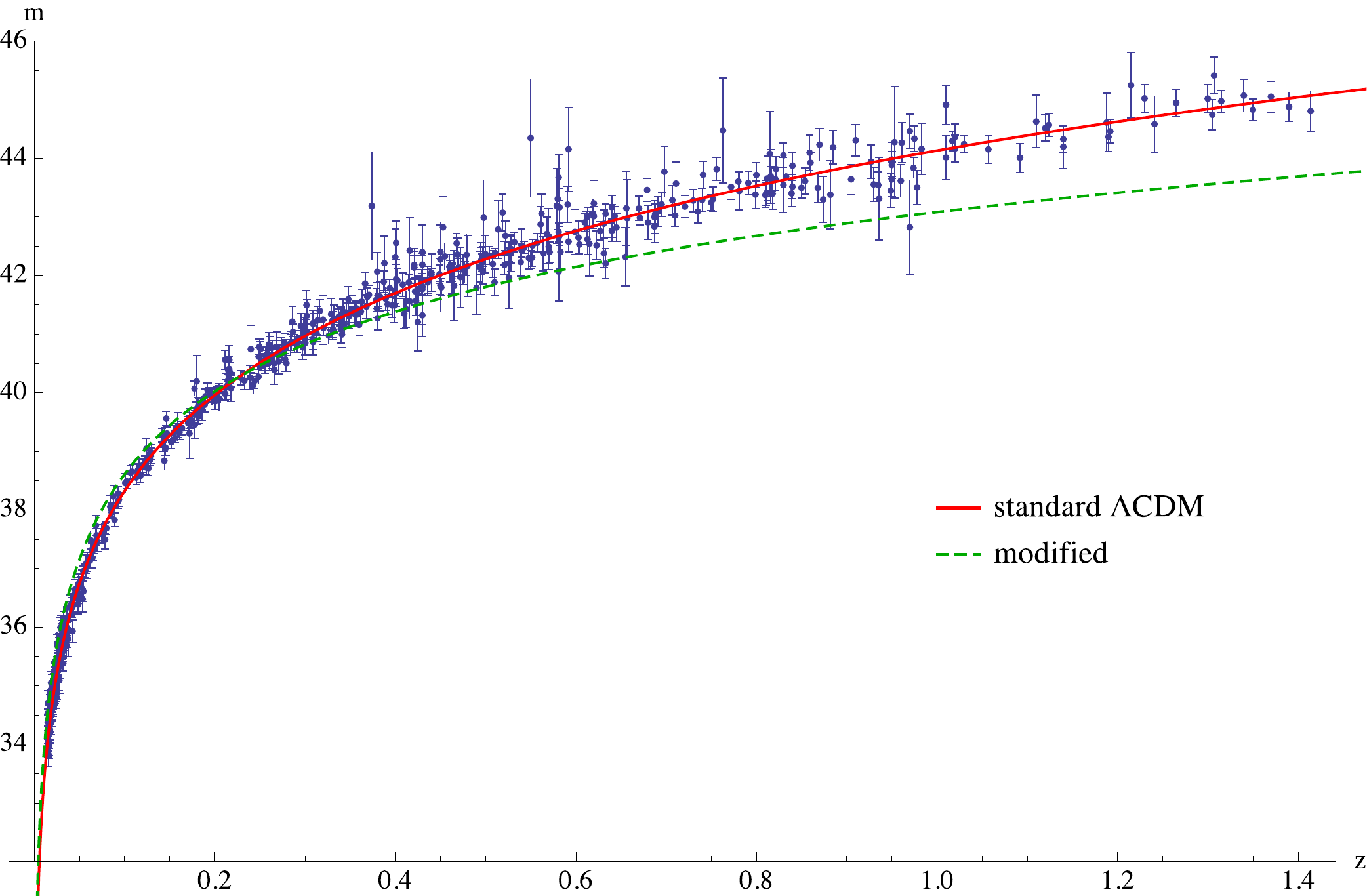}
	\caption{Magnitudes of the Union 2.1 SNe as function of redshift, together with the $ \Lambda $CDM best fit ($ \Omega_\Lambda = 0.72 $). The dashed green line is the best fit obtained for the modified Friedmann equations \eqref{eq:mod_Fried_coul_0} with $ p_\phi = 0 $ and two fluid components ($ w=0 $ and $ w=-1 $).}
	\label{fig:SNFit}
\end{figure}

Therefore, the closest one can get to the standard 4D evolution is in the limit $ r_c \to \infty $, in which case the behavior of $ \re^{a_0}$ is practically insensitive to the parameters $ \bar\rho_\mathrm{dust} $ and $ \bar\rho_\mathrm{cc} $. However, even in this limit the evolution is very different from 4D $ \Lambda $CDM. A corresponding supernova-fit\footnote{The fit does not converge but tends towards $ r_c \to \infty $, thus becoming insensitive to $ \bar\rho_\mathrm{dust} $ and $ \bar\rho_\mathrm{cc} $, and approaching the form shown in the plot.}  to the Union 2.1 data set \cite{SCP} is shown in Figure~\ref{fig:SNFit}, and clearly rules out the $ p_\phi = 0 $ model.

\subsection{Comment on Weinberg's no-go result}

In \cite{Weinberg:1988cp} it is argued that a dynamical adjustment mechanism which relies on introducing a finite number of additional fields that cancel the cosmological constant (CC) down to a small value cannot be implemented without invoking another fine tuning. The theorem relies on having four dimensional Einstein gravity. This assumption can be relaxed by embedding our world in a higher dimensional bulk equipped with higher dimensional Einstein gravity. Now the CC plays the role of a 4D maximally symmetric source localized on a brane which breaks translational invariance in the extra space directions. This symmetry pattern allows in principle for solutions where the CC is completely absorbed into extrinsic curvature and does not curve the brane geometry. It looks as if there is no need to further worry about Weinberg's argument. However, there is a potential pitfall: In the case of compact extra dimensions, we can consider the low energy effective description of the theory by simply integrating out the extra dimensions. This results in 4D Einstein gravity supplemented by a finite number of Kaluza Klein fields with masses below the EFT cutoff. Therefore, it should be possible to consistently understand the higher dimensional degravitation in this low energy theory in terms of a cancellation mechanism. This clearly contradicts Weinberg's theorem.

However, the situation is different for infinite extra dimensions. This theory is intrinsically higher dimensional and integrating out the extra dimensions results in a non-local theory of gravity to which the theorem cannot be applied~\cite{Dvali:2002pe,ArkaniHamed:2002fu}. In that context it is instructive to consider the modified Friedmann equation \eqref{eq:mod_Fried1_coul_0}. Removing the infinite dimension implies that the left hand side of this equation has to be set to zero. Then, by demanding $H_a=0$, the equation implies that also $\rho$ has to be tuned to zero. Thus, there is no dynamical adjustment or, equivalently, degravitation at work. Once we restore the infinite dimension, equation \eqref{eq:mod_Fried1_coul_0} admits a solution with an arbitrary value of $\rho$ and vanishing $H_a$ curvature. In accordance with the degravitation idea, the whole effect of $\rho$ is absorbed by the temporal curvature (i.e.\ expansion) in the compact extra dimension measured by $H_c$. This makes this mechanism especially interesting because it is different to the one discussed in \cite{Burgess:2011va} or \cite{Niedermann:2014bqa} where the effect of $ \rho $ is to curve the transverse spatial dimensions into a cone. Note, however, that it is not clear whether the choice $p_{\phi}=0$, which was made for the degravitating solutions, corresponds to another fine-tuning. To be more specific, it is not clear if there is a dynamical adjustment mechanism ensuring a vanishing azimuthal pressure which does not rely on a fine-tuning. This remains to be investigated.

\section{Conclusion and outlook}
\label{sec:conclusion}
In this work the DGP model was generalized by introducing an additional compact brane dimension. Assuming 4D FRW symmetries on the brane, the cosmology of this setup was investigated. Subsequently, a closed system of modified Friedmann equations describing the brane curvature evolution was derived. This was achieved by excluding any incoming gravitational waves as required by a source-free bulk.  To that end, an ``outgoing wave criterion'' based on a certain decomposition of the Weyl tensor was employed. 

Two physically distinct realizations of this setup were considered:
\begin{itemize}
\item {\it The compact brane direction is stabilized by introducing a non-trivial pressure $ p_\phi $ in angular direction.} The corresponding cosmological solutions are equivalent to the DGP solutions and correspond to a static bulk geometry. 

\item {\it The brane is allowed to freely expand or collapse in angular direction by setting $ p_\phi = 0 $.} In this case the brane generically acts as an emitter of 1D gravitational waves. Moreover, these solutions dynamically degravitate a 4D cosmological constant at the full non-linear level.
\end{itemize}

Subsequently, the phenomenological viability of the model was discussed. The first class of solutions has the same phenomenology as the DGP model as long as the compact dimension is stabilized. In particular, there is a standard 4D Friedmann regime which makes these solutions phenomenologically interesting. As a future direction, it might be interesting to resolve the stabilization mechanism dynamically by introducing additional fields. This could be achieved along the lines of \cite{Kaloper:2007ap} where a Scherk-Schwarz-like mechanism was employed to stabilize the compact dimension. From our analysis it can already be anticipated that such a stabilization should break down for a tension dominated stress-energy on the brane. This implies corrections to the DGP and standard Friedmann predictions at late times which would give rise to a new phenomenology.

The second class is ruled out due to the non-trivial angular dynamics, incompatible with a 4D regime, as is clearly inferred from Supernova observations. Therefore, the significance of these solutions is of purely conceptual nature as they provide an analytically realization of the degravitation mechanism at a non-linear level. In that context, it would be interesting to further investigate the linearized theory according to which the effective 4D graviton is a resonance, i.e.\ an infinite superposition of massive graviton states. This approach might illustrate how the degravitation mechanism can be incorporated in the 4D theory of a resonant graviton in a consistent and dynamical way. Moreover, it should be checked whether the choice $p_{\phi}=0$ can be achieved in a technically natural way.

\appendix

\section{Full bulk solution}
\label{ap:full_bulk_sol}

We will now show that under the assumption of vanishing Newton terms, the full bulk solution can be derived explicitly. To this end, it is useful to choose new coordinates\footnote{For the sake of notation we will still denote them by $ x^A = (t, y, x^i, \phi) $.} in which the metric takes the form
\begin{equation}
\label{eq:met_6D_new_coords}
	\rd s^2 = \re^{2(\eta - 3\alpha)} \left( -\rd t^2 + \rd y^2 \right) + \re^{2\alpha} \delta_{ij} \rd x^i \rd x^j + \re^{-6\alpha} w^2 \rd\phi^2,
\end{equation}
where $ \eta, \alpha $ and $ w $ are functions of $ (t, y) $. This is the generalization of the 4D coordinates often used (see e.g.\ \citep[chap. 22]{Stephani}) to describe geometries with azimuthal $ \phi $ symmetry and symmetry along a $ z $-direction perpendicular to $ \phi $, with the three $ x $-directions taking the role of $ z $. The numerical factors in \eqref{eq:met_6D_new_coords} were adapted such that the (modified) Einstein field equations, as in 4D, take a rather simple form\footnote{Since the brane is located at a constant coordinate position, and its surrounding is covered by a single coordinate patch, we can again use $ \delta $-functions to effectively implement Israel's junction conditions.}:
\newcommand{\tT}[2]{\mathcal{T}^{#1}_{\phantom{#1}#2}}
\begin{subequations}
\label{eq:efe_new_coords}
	\begin{align}
		\frac{w''}{w} - \frac{\ddot w}{w} &=  \tT{t}{t} + \tT{y}{y}, \\
		\alpha'' - \ddot \alpha +  \frac{w'}{w} \alpha' - \frac{\dot w}{w} \dot \alpha &= 4 \left( \tT{t}{t} + \tT{y}{y} - \tT{x}{x} + \tT{\phi}{\phi} \right), \\
		6 \left( {\alpha'}^2 + \dot \alpha^2 \right) -  \frac{w'}{w} \eta' - \frac{\dot w}{w} \dot \eta + \frac{\ddot w}{w} &= - \tT{y}{y}, \\
		6 \left( {\alpha'}^2 - \dot \alpha^2 \right) + \eta'' - \ddot \eta  &= \tT{\phi}{\phi}, \\
		12 \alpha' \dot \alpha -  \frac{w'}{w} \dot \eta - \frac{\dot w}{w} \eta' + \frac{\dot w'}{w} &= \tT{t}{y}.
	\end{align}
\end{subequations}
Here we defined
\begin{equation}
	\tT{A}{B} := \frac{\re^{2(\eta - 3\alpha)}}{M_6^4}\, \tilde T^A_{\phantom{A}B},
\end{equation}
with $ \tilde T^A_{\phantom{A}B} $ being the 6D energy momentum tensor, including the brane induced gravity terms:
\begin{subequations}
\begin{align}
	\tilde T^A_{\phantom{A}B} &= \left( T^{a}_{\phantom{a}b} - M_5^3 G^{(5)a}_{\phantom{(5)a}b} \right) \frac{\delta(y)}{\re^{\eta - 3\alpha}}  \, \delta^A_a \delta^b_B\\
	&\equiv \mathrm{diag} \left( -\tilde \rho, 0, \tilde p, \tilde p, \tilde p, \tilde p_\phi \right) \frac{\delta(y)}{\re^{\eta - 3\alpha}} \,.
\end{align}
\end{subequations}

The Newton terms (for $ y \neq 0 $) in these coordinates, after some simplifications using~\eqref{eq:efe_new_coords}, become
\begin{subequations}
\begin{align}
	\Phi^{(x)} & = \re^{2(3\alpha - \eta)} \frac{1}{2} \left[ \alpha'' - \ddot \alpha + {\alpha'}^2 - \dot \alpha^2 \right], \\
	\Phi^{(\phi)} & = \re^{2(3\alpha - \eta)} \frac{3}{2} \left[ \alpha'' - \ddot \alpha + 3\left( {\alpha'}^2 - \dot \alpha^2\right) \right],
\end{align}
\end{subequations}
and setting them equal to zero is equivalent to
\begin{equation}
\label{eq:alpha_coul_0}
	{\alpha'}^2 = \dot \alpha^2 \quad \Leftrightarrow \quad \alpha' = \pm \dot \alpha.
\end{equation}
The bulk ($ y \neq 0 $) Einstein equations \eqref{eq:efe_new_coords} then reduce to
\begin{subequations}
\begin{gather}
\label{eq:w_coul_0}
	w' = \pm \dot w \\
\label{eq:eta1_coul_0}
	\eta'' - \ddot \eta = 0 \\
\label{eq:eta2_coul_0}
	12 \dot \alpha^2 + \frac{\ddot w}{w} - \frac{\dot w}{w} \left( \dot \eta \pm \eta' \right) = 0,
\end{gather}
\end{subequations}
where the choice of signs in \eqref{eq:w_coul_0} and \eqref{eq:eta2_coul_0} has to be the same as for $ \alpha $ in \eqref{eq:alpha_coul_0}. So $ \alpha $ and $ w $ are both functions of $ (t \pm y) $, i.e.\ they are both either \emph{left-moving 1D waves}, or both \emph{right-moving 1D waves}. $ \eta $ can in general be a superposition of left- and right-moving 1D waves.
In order to create the $ \delta $-sources on the right hand side of \eqref{eq:efe_new_coords}, the following junction conditions have to be fulfilled:
\begin{subequations}
\label{eq:junction_cond_new_coord}
\begin{align}
	\jump{\alpha'} &= - \frac{\re^{\eta_0 - 3\alpha_0}}{4 M_6^4} \left( \tilde \rho + \tilde p - \tilde p_\phi \right) \\
	\frac{\jump{w'}}{w_0} &= -\frac{\re^{\eta_0 - 3\alpha_0}}{M_6^4} \tilde \rho, \\
	\jump{\eta'} &= \frac{\re^{\eta_0 - 3\alpha_0}}{M_6^4} \tilde p_\phi.
\end{align}
\end{subequations}
Now the only way to satisfy these, in the nontrivial case of non-vanishing right hand sides, is to choose $ \alpha $ (and thus also $ w $) to be of opposite wave character (left-/right-moving) on the left and right side of the brane at $ y = 0 $.
Since we do not want waves propagating towards the brane, we will take them to be right-moving on the right, and left-moving on the left; so we choose the plus sign for $ y < 0 $, and the minus sign for $ y>0 $.
Requiring $ \alpha $ and $ w $ to be continuous then yields:
\begin{subequations}
\label{eq:ans_coul_0}
\begin{align}
\label{eq:alpha_ans_coul_0}
	\alpha(t, y) & = \alpha_0(t - |y|) \\
\label{eq:w_ans_coul_0}
	w(t, y) & = w_0(t - |y|) \\
\label{eq:eta_ans_coul_0}
	\eta(t, y) & = 
	\begin{cases}
		\eta_{\text{L}<}(t + y) + \eta_{\text{R}<}(t - y) & (y < 0) \\
		\eta_{\text{L}>}(t + y) + \eta_{\text{R}>}(t - y) & (y > 0)
	\end{cases}
\end{align}
\end{subequations}
The only remaining Einstein equation \eqref{eq:eta2_coul_0} (for $ \dot \alpha_0 \neq 0 $), as well as continuity of $ \eta $ further imply\footnote{Note that now all the metric functions have a reflection symmetry around $ y=0 $ --- which we didn't assume in this derivation --- in accordance with the aforementioned result that zero Newton-like field components (and nonzero $ \tilde\rho, \tilde p $) imply that the mean-values of the first $ r $-derivatives vanish.}
\begin{equation}
\label{eq:eta_in_out}
	\eta_{\text{L}<} = \eta_{\text{R}>} + C, \qquad \eta_{\text{L}>} = \eta_{\text{R}<} + C,
\end{equation}
with some constant $ C $, allowing to eliminate two of the four functions appearing in \eqref{eq:eta_ans_coul_0}, say $ \eta_{\text{L}<} $ and $ \eta_{\text{L}>} $.
The jumps of the first derivatives of the metric functions can now be expressed directly in terms of on-brane functions, which --- together with equation \eqref{eq:eta2_coul_0} --- yields
\begin{subequations}
\begin{gather}
\begin{align}
	\jump{\alpha'} & = -2 \dot \alpha_0 \\
	\jump{w'} & = -2 \dot w_0 \\
	\jump{\eta'} & = -2 \left( \dot \eta_{\text{R}>} - \dot \eta_{\text{R}<} \right)
\end{align}\\
	12 \dot \alpha_0^2 + \frac{\ddot w_0}{w_0} - 2 \frac{\dot w_0}{w_0} \dot \eta_{\text{R}>} = 0.
\end{gather}
\end{subequations}
Since the three jumps can be expressed in terms of the on-brane sources using the junction conditions \eqref{eq:junction_cond_new_coord}, this constitutes a system of four ODEs for the five unknown functions\footnote{Note that $ \eta_0 $ can be written as $ \eta_0(t) = \eta_{\text{R}<}(t) + \eta_{\text{R}>}(t) + C $, and we assume some equations of state to be given for the both pressure components $ p $ and $ p_\phi $.} $ \alpha_0(t), w_0(t), \eta_{\text{R}>}(t), \eta_{\text{R}<}(t) $ and $ \rho(t) $. But there is still a residual gauge freedom which allows us to require
\begin{equation}
\label{eq:gauge_new_coords}
	\eta_0(t) = 3\alpha_0(t)
\end{equation}
by an appropriate redefinition of the $ t $ and $ r $ coordinates (see Appendix \ref{ap:fix_gauge}), so that the five dimensional induced metric becomes
\begin{equation}
\label{eq:ind_metric_5D}
	\rd s^2_{(5)} = -\rd t^2 + \re^{2\alpha_0} \delta_{ij} \rd x^i \rd x^j + \re^{-6\alpha_0} w_0^2 \rd\phi^2 \,.
\end{equation}
This allows us to eliminate, say $ \eta_{\text{R}<} $, thus leading to a closed system, which can be brought into the form:
\begin{subequations}
\label{eq:mod_Fried_new_coords}
\begin{align}
	2 \dot \alpha_0 + \jump{\alpha'} & = 0 \,, \\
	2 \dot w_0 + \jump{w'} & = 0 \,, \\
	\dot{\jump{w'}} +12 w_0 \dot\alpha_0 \jump{\alpha'} - \dot w_0\left( 3 \jump{\alpha'} + \jump{\eta'} \right) & =0 \,, \\
\label{eq:eta_out_coul_0}
	4 \dot \eta_{\text{R}>} + 3\jump{\alpha'} + \jump{\eta'} & = 0 \,.
\end{align}
\end{subequations}
The gauge \eqref{eq:gauge_new_coords} is particularly useful, because it renders the induced 5D metric independent of $ \eta_0 $. Therefore, the brane induced gravity terms contributing to $ \jump{\alpha'}, \jump{w'} $ and $ \jump{\eta'} $ also become independent of $ \eta_0 $, and so the first three equations of \eqref{eq:mod_Fried_new_coords} form a closed system for the three functions $ \alpha_0(t), w_0(t) $ and $ \rho(t) $. After making the identifications
\begin{equation}
	a_0 = \alpha_0, \qquad c_0 = \log(w_0) - 3\alpha_0 \,,
\end{equation}
following from \eqref{eq:ind_metric_5D}, as well as 
\begin{align}
	\jump{\alpha'} & = \jump{a'}, & \frac{\jump{w'}}{w_0} & = 3\jump{a'} + \jump{c'}, & \jump{\eta'} & = 3\jump{a'} + \jump{n'},
\end{align}
which follow from comparing the junction conditions in the two different sets of coordinates, i.e.\ equations \eqref{junction_cond_solved} and \eqref{eq:junction_cond_new_coord}, one can easily verify that they reproduce the set of equations~\eqref{eq:mod_Fried_coul_0} derived earlier. To be more precise, one recovers the $ + $ branch of this solution. The $ - $ branch corresponds to waves that are traveling onto the brane.

The time evolution of $ \eta_{\text{R}>}(t) $ decouples, and it is obtained by simply integrating \eqref{eq:eta_out_coul_0}. The full function $ \eta(t, y) $ can then be written as:
\begin{equation}
\label{eq:eta_sol_coul_0}
	\eta(t, y) = 3 \alpha_0(t + |y|) - \eta_{\text{R}>}(t + |y|) + \eta_{\text{R}>}(t - |y|)
\end{equation}
This completes the derivation of the complete bulk geometry, which will be known explicitly once the on-brane system \eqref{eq:mod_Fried_new_coords}, or equivalently \eqref{eq:mod_Fried_coul_0}, is solved for specific equations of state for the on-brane matter content.

Now that we know the whole bulk geometry, its interpretation becomes quite obvious: All the metric functions are purely 1D waves. In principle these could be mere coordinate artifacts, but the wave components of the Weyl tensor are given by:
\begin{subequations}
\begin{align}
	\Oout(t, y) & = 
	\begin{cases}
		0 & (y < 0) \\
		2 \left[ \ddot\alpha_0(t - y) + 7 \dot\alpha_0(t - y)^2 - 2 \dot\alpha_0(t - y) \dot\eta_{\text{out}>}(t - y) \right] & (y > 0)
	\end{cases}\\
	\Oin(t, y) & = 
	\begin{cases}
		2 \left[ \ddot\alpha_0(t + y) + 7 \dot\alpha_0(t + y)^2 - 2 \dot\alpha_0(t + y) \dot\eta_{\text{out}>}(t + y) \right] & (y < 0) \\
		0 & (y > 0)
	\end{cases}
\end{align}
\end{subequations}
These are generally non-zero, showing that there are in fact real physical waves: a bulk observer would see test particles accelerated as described in section~\ref{sec:interpret_weyl}. Moreover, they are indeed propagating away from the brane on both sides, as was intended by the choice~\eqref{eq:alpha_ans_coul_0}, \eqref{eq:w_ans_coul_0} --- despite the fact that the solution \eqref{eq:eta_sol_coul_0} of $ \eta $ is a superposition of incoming and outgoing waves, so this part of the metric is indeed a pure coordinate artifact.

\section{Fixing the residual gauge}
\label{ap:fix_gauge}

In this section we will show that the gauge choice \eqref{eq:gauge_new_coords} is always possible. To this end, note that the form of the metric \eqref{eq:met_6D_new_coords} is unchanged under a redefinition of the $ t $ and $ y $ coordinates of the form
\begin{equation}
\label{eq:coord_trf}
	\begin{pmatrix}
	t\\
	y
	\end{pmatrix}
	\mapsto
	\begin{pmatrix}
	\bar t\\
	\bar y
	\end{pmatrix}
	=
	\begin{pmatrix}
	f(t, y)\\
	g(t, y)
	\end{pmatrix}
\end{equation}
with
\begin{align}
	&\begin{pmatrix}
	\dot f\\
	f'
	\end{pmatrix}
	= \pm
	\begin{pmatrix}
	g'\\
	\dot g
	\end{pmatrix}
	&&\text{and}&
	{g'}^2 \neq {\dot g}^2.
\end{align}
After this transformation, the metric functions will take the form
\begin{align}
	& \bar \alpha = \alpha, && \bar w = w, && \re^{-2\bar\eta} = \re^{-2\eta} \left( {g'}^2 - {\dot g}^2 \right).
\end{align}
If we now define the function
\begin{equation}
	h(t) := \int^t \re^{\eta_0(t') - 3 \alpha_0(t')}\,\rd t',
\end{equation}
then a transformation of the form \eqref{eq:coord_trf} with
\begin{subequations}
\begin{align}
	f(t, y) & = \frac{1}{2} \bigl[ h(t + y) + h(t - y) \bigr] \\
	g(t, r) & =	\frac{1}{2} \bigl[ h(t + y) - h(t - y) \bigr]
\end{align}
\end{subequations}
will lead to $ \bar \eta_0 = 3 \alpha_0 $.

\acknowledgments
The authors would like to thank Felix Berkhahn and Stefan Hofmann for inspiring discussions.
The work of FN and RS was supported by the DFG cluster of excellence `Origin and Structure of the Universe'.

\newpage
\bibliographystyle{utphys}
\bibliography{cosmo_ring_v2}

\end{document}